\documentclass[print]{raa}
\usepackage{graphicx,times}
\usepackage{subfigure}
\usepackage{natbib}
\usepackage{amssymb,amsmath}
\usepackage{pdfpages}
\bibpunct{(}{)}{;}{a}{}{,}

\def\aap{$A\&A$}
\def\mnras{$MNRAS$}
\def\aj{$AJ$}
\def\apj{$ApJ$}
\def\apjl{$ApJL$}
\def\apjs{$ApJS$}
\def\pasp{$PASP$}

\usepackage[a4paper=true,dvipdfm=true,pagebackref=true]{hyperref}
\hypersetup{pdftitle = The title of my PDF, pdfauthor = My name, pdfsubject= The subject, pdfkeywords = keyword1 keyword2 keyword3}
\hypersetup{colorlinks = true, linkcolor = green, anchorcolor = red, citecolor = blue, filecolor = red, pagecolor = red, urlcolor = red}

\begin{document}

   \title{Powerful CMD: A Tool for Colour-Magnitude Diagram Studies
$^*$
\footnotetext{\small $*$ Supported by the National Natural Science Foundation of China (No. 11563002)
and Yunnan Science Foundation (application No. C0120150619).}
}

 \volnopage{ {\bf 2012} Vol.\ {\bf X} No. {\bf XX}, 000--000}
   \setcounter{page}{1}

   \author{Zhong-Mu Li\inst{1}, Cai-Yan Mao\inst{1}, Qi-Ping Luo
      \inst{1},  Zhou Fan\inst{2}, Wen-Chang Zhao\inst{1}, Li Chen\inst{1}, Ru-Xi Li\inst{1}, Jian-Po Guo\inst{3}
   }

   \institute{$^{1}$ Institute for Astronomy and History of Science and
Technology, Dali University, Dali 671003, China; {\it zhongmuli@126.com}\\
    $^{2}$ Key Laboratory for Optical Astronomy, National Astronomical Observatories, Chinese Academy of
Sciences, A20, Datun Road, Chaoyang District, Beijing 100012, China\\
    $^{3}$ College of science and technology, Puer University, Puer 665000, China\\
\vs \no
   {\small Received 2012 June 12; accepted 2012 July 27}
}

\abstract{We present a new tool for colour-magnitude diagram (CMD) studies, $Powerful~CMD$.
This tool is built on the basis of the advanced stellar population synthesis (ASPS) model,
in which single stars, binary stars, rotating stars, and star formation history have been taken into account.
Via $Powerful~CMD$, the distance modulus, colour excess, metallicity, age, binary fraction, rotating star fraction,
and star formation history of star clusters can be determined simultaneously from observed CMDs.
The new tool is tested via both simulated and real star clusters.
Five parameters of clusters NGC6362, NGC6652, NGC6838 and M67 are determined and compared to other works.
It is shown that this tool is useful for CMD studies,
in particular for those with the data of the \emph{Hubble Space Telescope (HST)}.
Moreover, we find that the inclusion of binaries in theoretical stellar population models may lead to
smaller colour excess compared to the case of single star population models.
\keywords{(stars:) Hertzsprung-Russell and C-M diagrams; (Galaxy:) globular clusters: general;galaxies: clusters: general
}
}

   \authorrunning{Z.-M. Li et al. }            
   \titlerunning{Powerful Tool for CMDs}  
   \maketitle

%
\section{Introduction}           
\label{sect:intro}

Colour-magnitude diagram (CMD) is the observed counterpart of Hertzsprung-Russell diagram.
It shows the distribution of stars in the magnitude versus colour plane.
CMDs are of key importance for the studies of star clusters.
Many astrophysical properties, e.g., distance modulus, colour excess, metallicity, age, binary fraction
and star formation history, can be determined from CMDs.
Such results can be widely used for studying the evolution of stars, star clusters and galaxies.
The age of the universe can also be constrained by the age of the oldest globular clusters, which can be derived from CMDs.
The CMD of a star cluster can be obtained by observing stars within the cluster in two passbands.
There have been many observations for the CMDs of star clusters.
Both ground-based and space telescopes have been used.
In particular, a quantity of high-quality CMDs are obtained by the \emph{Hubble Space Telescope} (\emph{HST}).
This makes it possible for us to study star clusters in detail via CMDs.
One can read many papers about CMD studies, e.g., \cite{1996Fusi}, \cite{1998Olsen}, \cite{2006Mieske}, \cite{2007Mackey}, \cite{2013VandenBerg}, \cite{Yang2013}, \cite{2012ApJ...761L..22L,2015ApJ...802...44L}, \cite{2015ApJ...807...25B}, and \cite{2016A&A...586A.148N}.

One method that is widely used to study star clusters is based on the comparison between the observed and synthetic CMDs.
In most works, the synthetic CMDs are described by some isochrones (distribution of a population of stars with the same metallicity and age but various masses).
The comparison of observed and theoretical CMDs can be simply done \emph{by eye}, but this is usually for a small set of isochrones.
This method is unsuitable for studying a large amount of clusters. It is not suitable for some detailed studies,
which need numerous isochrones, either.
Therefore, many works tried to fit CMDs via some statistical methods, for example: \begin{itemize}
                                                                        \item 1) Bayesian method;
                                                                        \item 2) $\chi^{2}$ method;
                                                                        \item 3) $\tau^{2}$ method;
                                                                        \item 4) likelihood statistic.
                                                                      \end{itemize}
Bayesian method was used by e.g., \cite{2006ApJ...645.1436V} and \cite{2009ApJ...696...12D}. This technique is to use information from the data and from our prior knowledge to obtain posterior distributions on the parameters (e.g., metallicity, age, and initial mass of cluster stars) of stellar population models.
$\chi^{2}$ method was used by e.g., \cite{2001ApJS..136...25H}, \cite{2005A&A...435...77K}, and \cite{2006A&A...454..511C}. This method is good for the ideal case with single Gaussian uncertainties but it has been widely used for similar cases.
$\tau^{2}$ method was used by \cite{2006MNRAS.373.1251N} and \cite{2010ApJ...723..166D}. A parameter, $\tau^{2}$ = -2 ln\emph{P} (where \emph{P} is probability), is used to find the best-fit models. The distribution of $\tau^{2}$ is different to $\chi^{2}$ when uncertainties are small, but the two goodness parameters have similar distributions for a large uncertainty case.
A likelihood statistic was used by \cite{2002A&A...390..121K}, and $\chi_{\rm e}^{2}$ method was used by e.g., \cite{2002MNRAS.332...91D}. The likelihood \emph{L} is given by $L = \prod P_{\rm i}$, where $P_{\rm i}$ is the model probability function evaluated at the CMD position of the $i$th star.
Such methods are much more quantitative than the \emph{by eye} method, and they improve the reliability of CMD fitting obviously.
A limitation of the previous works is that they are based on some classical stellar population synthesis models,
and some of them took only a part of CMD, e.g., main sequence, rather than the whole CMD.
In addition, no tool is able to simultaneously determine cluster properties including e.g.,
binary fraction, rotating star fraction, and star formation history.
This makes it difficult to get a comprehensive understanding about star clusters from CMDs.

This work introduces a tool, i.e., $Powerful~CMD$, which is aiming for determining the distance modulus, colour excess, metallicity, age, binary fraction
and star formation history from CMDs, which is produced by Dr. Zhongmu Li.
It can also be used for building the CMDs of various kinds of stellar populations.
The work principle, tests and application of the new tool are presented by this paper.

This paper is organized as follows. In Sect. 2, we briefly introduce the stellar population model used by $Powerful~CMD$.
In Sect. 3, we introduce how $Powerful~CMD$ builds synthetic CMDs and finds the best-fit models. Next, in
Sect. 4, we present some tests to the new tool. In Sect. 5, we apply the tool to a few star clusters.
Finally, we give our discussions and conclusions in Sect. 6.

\section{Stellar population model}

$Powerful~CMD$ works on the basis of the advanced stellar population synthesis (ASPS) model of Dr. Zhongmu Li.
The ASPS model was developed from the rapid stellar population synthesis model (RPS)
\citep{2008MNRAS.387..105L, 2008ApJ...685..225L, 2010RAA....10..135L, 2011ASPC..451...51Z, 2012MNRAS.424..874L, 2013ApJ...776...37L, 2015ApJ...802...44L}.
A feature of ASPS is taking the effects of binary stars and rotating stars into account \citep{2012ApJ...761L..22L, 2015ApJ...802...44L, 2016arXiv1604.07156}.
The current version of ASPS takes the initial mass function (IMF) of \citet{1955ApJ...121..161S} (Salpeter IMF),
8 metallicities (0.0001, 0.0003, 0.001, 0.004, 0.008, 0.01, 0.02 and 0.03), 151 ages (0--15\,Gyr with an interval of 0.1\,Gyr),
and 7 rotating star fractions (0, 0.1, 0.3, 0.5, 0.7, 1.0 for stars following the rotation rate distribution of \cite{2007A&A...463..671R},
and a gaussian distribution with mean and deviation of 0.7 and 0.1 for all stars).
Although there is still some uncertainty in the IMF for stars less massive than the Sun,
it does not affect the results obviously, because usually bight stars (i.e., massive stars at zero age) are used for CMD studies.
The results from Geneva code are used for considering stellar rotation in ASPS, and our results are consistent with some recent works, e.g., \cite{2015ApJ...807...25B}, \cite{2015MNRAS.453.2637D} and \cite{2015MNRAS.453.2070N}. Because the Geneva models do not include low-mass ($<$ 1.7 solar mass) rotating stars,
rotation is not considered by ASPS for star clusters older than about 2\,Gyrs.
In fact, rotation affects old clusters ($>$ 2\,Gyrs) much more slightly compared to the young ones, as such clusters are dominated by low-mass and slow-rotating stars.
The basic stellar population models of ASPS take a binary fraction of 50\%.
Note that binaries here mean those with orbital period less than 100\,yr.
They are different from interacting binaries.
Therefore, the binary fraction in this paper is usually larger than other works.
Because every single and binary star in basic models can be removed, binary fraction can be changed to any value between 0 and 1 in studies.
The star sample of ASPS is generated by a Monte Carlo technique following the Salpeter IMF at zero age, then
all stars are evolved to present day using the rapid stellar evolution code of
\citet{1998MNRAS.300..977H} and \citet{2002MNRAS.329..897H}.
The effect of rotation on the evolution of stars are calculated using the results of \cite{2013A&A...553A..24G},
via taking various rotation rate distribution, i.e., the above-mentioned seven rotating star fractions.
Readers are kindly invited to read another paper, \cite{2016arXiv1604.07156}, to learn more details about ASPS.

\section{Synthetic CMDs and CMD fitting}
\subsection{Synthetic CMDs}

ASPS model only supplies some basic stellar populations with 50 per cent binaries,
but we can change the binary fraction of a stellar population in the synthesis of CMDs because the binarity (binary or single) of every star has been marked.
This is easily achieved by removing some random binaries or single stars from the basic models.
Moreover, stellar population models with various rotating star fractions can be chosen from the basic models of ASPS.
The CMDs of simple stellar populations (SSPs) with a fraction of binaries and a fraction of rotating stars can be built in this way (e.g., panels (a), (b), (e) and (f) of Fig. 1).
When we build the CMDs of composite stellar populations (CSPs), $Powerful~CMD$ puts a few SSPs together.
Because the stars of most studied star clusters ($\sim$ 80 per cent) have no obvious metallicity difference,
we assume that a CSP consists of stars with the same metallicity but various ages.
Here we do not take into account chemical evolution,
because observations did not show obvious metallicity difference for the stars of a not too old cluster.
The star number of each SSP is assigned according to a chosen star formation history, for clearly.
In this way, the intrinsic CMDs of SSPs and CSPs are built. Fig. 1 shows some examples.
Each panel presents the CMD of a kind of stellar population.
In detail, panels (a)--(d) contain no rotating stars, while the others include a fraction of rotators.
Panels (a), (c), (e) and (g) contain no binary stars, while the others contain some binaries.
Panels (c), (d), (g) and (h) are for CSPs and the others for SSPs.
The model inputs of these example stellar populations are listed in Table 1.

In fact, the observed CMDs are obviously affected by distance, colour excess, and uncertainties in magnitudes and colours.
This makes observed CMDs different from the intrinsic CMDs of theoretical stellar populations.
Therefore, it is necessary to include this effect in synthetic CMDs.
In detail, the magnitudes of stars are moved toward less luminous direction by adding a distance modulus,
and their colours are moved toward redder direction by adding a colour excess.
Because colour excess corresponds to a faintness of stars,
we need to correct distance modulus after getting the results.
A correlation between colour excess and magnitude change is needed for this operation,
which depends on the dust distribution model of the Milky Way.

A difficulty to model CMDs accurately is taking into account the observational uncertainties of magnitudes correctly.
Such uncertainties are caused mainly by equipment for observation, software and method for dealing with the data (i.e., photometry process), and random errors.
The uncertainties caused by equipment and photometry process dominate observational uncertainties.
Usually, the uncertainties caused by equipment are reported, but those caused by photometry process are not so clear.
In order to quantify the uncertainties due to photometry process,
we can make some artificial star tests (ASTs) (e.g., \citealt{1996Sandquist}, \citealt{2001ApJS..136...25H}, \citealt{2008Anderson} and \citealt{2010MNRAS.403.1156R}).
Some images consisting of many artificial stars with known magnitudes and colours are built and then processed with a photometry software to measure their magnitudes.
The magnitude uncertainties (AST uncertainties) caused by photometry process are given by comparing the input and measured magnitudes.
Meanwhile, the star completeness at any CMD region is given by comparing the input and measured star fractions.
Although ASTs can account only for a part of the observational errors, some tests (e.g., \citealt{2010MNRAS.403.1156R}) showed that AST technique can be used for some CMD works.
Fig. 2 shows the uncertainty versus magnitude relations of stars in two simulated star clusters (S1 and S6).
Such relations will be used for adding AST uncertainties to intrinsic CMDs of stellar populations.
When such uncertainties are considered,
the CMDs of stellar populations usually become significantly scattered and seem closer to the observed ones.
In the case that uncertainties due to photometry process are much larger than those caused by equipment,
we can use the results of ASTs as the final observational uncertainties.
However, we should note that the real observational errors are larger than AST ones.
For example, differential reddening and PSF variations also contribute to the observational errors (\citealt{milone2009,milone2012}).
We suggest to take these errors into account if possible.
After including observational uncertainties, CMDs in Fig. 1 become to the case of Fig. 3.
Note that many other methods can be used for estimating the observational uncertainties,
so one can choose his own methods.

\subsection{CMD fitting}
In order to make CMD fitting more convenient and effective,
$Powerful~CMD$ divides a CMD into many cells by taking fixed colour and magnitude intervals and counts the stars in each cell.
Then it fits the Hess diagrams (Hess 1924) instead of the original CMDs to find the best-fit parameters of star clusters.
As the standard case, a CMD plane is divided into 1500 cells, including 50 colour bins and 30 magnitude bins, but it can be changed.
It is suggested to test the effect of bin numbers on the result.
Our test shows that this selection is able to reproduce most of the input CMDs.
One can also take fixed intervals for colour and magnitude for the fitting.
In principle, if synthetic CMDs are well built, the result will not be affected too much by the colour and magnitude bins when they are larger than about 30.
The star fractions of observed and theoretical CMDs in the same cell are denoted by $f_{\rm ob}$ and $f_{\rm th}$ for comparison.

Although a few statistics can be used for finding the best-fit models,
they all have both advantages and disadvantages. Thereby, $Powerful~CMD$ uses three statistics to identify the best-fit model.
Users can choose the statistic for CMD fitting. The three statistics in the current version include the widely used $\chi^{2}$, effective $\chi_{e}^{2}$ \citep{2003AJ....125..770B}, and weighted average difference ($WAD$, \citealt{2015ApJ...802...44L}). Note that the $\chi_{e}^{2}$ statistic is proper for dealing with Poission-distributed data.
In \cite{2015ApJ...802...44L}, we compared WAD with $\chi^{2}$ and $\chi_{e}^{2}$ statistics.
It has shown that $WAD$ is a good indicator for determining the best-fit parameters of star clusters, and it gives similar results to $\chi^{2}$.
In detail, $WAD$, $\chi^{2}$ and $\chi_{e}^{2}$ are calculated via equations 1--4.
\begin{equation}\label{eq3}
    WAD = \frac{\Sigma [{\omega_i.|f_{ob}-f_{th}|}]}{\sum\omega_i}~,
\end{equation}
where $\omega_i$ is the weight of $i$th cell, and
$f_{ob}$ and $f_{th}$ are star fractions of observed and theoretical
CMDs in the same cell. Both $f_{ob}$ and $f_{th}$ are between 0 and 1.
$\omega_i$ is greater than 0, and it is calculated as
\begin{equation}\label{eq4}
    \omega_i = \frac{1}{|1-C_i|}~,
\end{equation}
where $C_i$ ($<$ 1) is the completeness of $i_{\rm th}$ cell, and it can be estimated via ASTs (see \citealt{2015ApJ...802...44L}).
Here $|1-C_i|$ gives the uncertainty of star fraction in the $i_{\rm th}$ cell.
\begin{equation}\label{eq4}
    \chi^2 = \Sigma \frac{(f_{ob}-f_{th})^2}{(1-C_i)^2},
\end{equation}

and
\begin{equation}\label{eq4}
    \chi_e^2 = 2\Sigma [(f_{ob}-f_{th})+f_{th}.log(f_{th}/f_{ob})].
\end{equation}

CMD fitting can be completed automatically by $Powerful~CMD$, when the observational data and control file have been prepared.
In the control file, one needs to set the number of stars in theoretical models, colour for fitting, whether considering observational errors and star incompleteness, CMD range for fitting, and the range of star formation mode. The ranges and steps for distance modulus, colour excess, metallicity, age, age spread, binary fraction, and rotator fraction are also needed to be given in this file. This makes the tool friendly to use.

$Powerful~CMD$ is possibly able to serve some hot researches, e.g., the extended main-sequence turn off of star clusters of 100--2000\,Myr, which has attracted much attention but the reason is still not clear. Both a spread of age (e.g., \citealt{2008MNRAS.424..874L, milone2009}) and a spread of stellar rotation rate (e.g., \citealt{2015MNRAS.453.2637D}) can possibly explain the observation . Aims to find whether $Powerful~CMD$ can be used for disentangling the effects of spread of age and spread of rotation rate of stars, we did a test. We found that $Powerful~CMD$ can disentangle the degeneracy between age spread and rotation rate spread partially. If star clusters exist extended or multiple red clumps, $Powerful~CMD$ usually prefers age spread as the reason of extended turn-off, because such special red clump structure is possibly not formed from rotation, as rotation of stars becomes much slower when they leave main sequence and loss lot of angular momentum. Similarly, for star clusters younger than 0.5\,Gyr, $Powerful~CMD$ usually reports age spread because rotation contributes to extended turn-off slightly (\citealt{2015ApJ...802...44L}). For other clusters, $Powerful~CMD$ can not give reliable conclusion on the reason of extended turn off, although a best-fit model can be given. Note that the results depends on stellar population model.

\section{Test to the new tool}
\subsection{Building CMDs of various stellar populations}
A function of $Powerful~CMD$ is to build CMDs based on different stellar population assumptions.
Different metallicities, ages, binary fractions, rotating star fractions, star formation histories,
distance moduli and colour excesses can be taken,
and CMDs can be generated quickly (about decades of seconds for a few thousand of stars but it depends on computer).
Fig. 1 has given some examples without observational uncertainties,
so we test the CMD building function by taking colour and magnitude uncertainties into account here.
As examples, Fig. 3 shows the CMDs of eight simulated star clusters (black points).
The model inputs are listed in Tables 2 and 3, while the corresponding magnitude uncertainties of stars of two clusters can be seen in Fig. 2.
We are shown that $Powerful~CMD$ has the ability to build the CMDs of various stellar populations.
This is helpful for many studies of star clusters and galaxies.

\subsection{Fitting CMDs of stellar populations}
The $WAD$ method was tested in a previous work \citep{2015ApJ...802...44L} and its reliability has been shown,
but this method cannot give a constraint on the uncertainties of cluster parameters,
because we do not know the distribution of $WAD$ values.
Thus we choose $\chi^{2}$ as the goodness indicator of fit and test $Powerful~CMD$ in this paper.
The results show that this method can recover most cluster parameters,
when we use some artificial star clusters to test $Powerful~CMD$.
Fig. 4 shows the process (only a few example steps) of CMD fitting.
We observe that this code is able to find the best-fit model step by step.
In detail, 51 simulated star clusters, including S1--S8, are used for testing the tool.
Fig. 5 gives the test results, in which the input and recovered values of seven parameters are compared.
We find that distance modulus, colour excess, metallicity, youngest-component age, age spread, binary fraction and rotator fraction are well recovered.
Note that age spread is described by another parameter, $N_{\rm sf}$, which means the number of star formations from the youngest component and with an age interval of 0.1\,Gyr.
When checking the star formation mode, the input modes of 12 are recovered correctly, within 15 simulated star clusters.
Therefore, $Powerful~CMD$ recovered the input parameters of most simulated star clusters.
For a quantitative comparison purpose, Tables 2 and 3 list the input and fitted parameters of eight simulated clusters (S1--S8).

\begin{table} 
 \caption{Input parameters for eight simulated star clusters in Fig. 1.
 $Z$, $f_{\rm b}$ and $f_{\rm rot}$ denote metallicity, binary fraction and rotator fraction, respectively.}
 \label{symbols}
 \begin{tabular}{ccccc}
  \hline\hline
No.	&$Z$	    &Age (Gyr)  &$f_{\rm b}$   &$f_{\rm rot}$         \\
 \hline
a	&	0.01	&	0.5	&	0	&	0	\\
b	&	0.001	&	1.5	&	0.5	&	0	\\
c	&	0.004	&	0.9,1.0,1.1,1.2	&	0	&	0	\\
d	&	0.02	&	0.6,0.7,0.8	&	0.5	&	0	\\
e	&	0.008	&	1.0	&	0	&	0.5	\\
f	&	0.008	&	1.0	&	0.5	&	0.5	\\
g	&	0.008	&	0.9,1.0,1.1,1.2,1.3	&	0	&	1.0	\\
h	&	0.008	&	0.9,1.0	&	0.7	&	1.0	\\
 \hline
 \end{tabular}
 \end{table}

\begin{table*} 
 \caption{Input and fitted parameters for eight simulated star clusters.
 $CE$ means color excess. Subscript ``in'' and ``fit'' denote input and fitted parameters.
 Parameter ranges are 1 $\sigma$ ranges.}
 \label{symbols}
 \begin{tabular}{cccccccccccc}
  \hline\hline
No.	&$(m-M)_{\rm in}$	&$(m-M)_{\rm fit}$	&range	        &$CE_{\rm in}$	&$CE_{\rm fit}$	&range      &$Z_{\rm input}$	 &$Z_{\rm fit}$	&range	     \\
    &[mag]	            &[mag]	            &[mag]	        &[mag]       	&[mag]       	&[mag]      &              	    &            	 &	         \\
 \hline
S1	&	18.40	&	18.37	&	18.28--18.58	&	0.08	&	0.08	&	0.04--0.12	&	0.0010	&	0.0010	&	 0.0003--0.0010	\\		
S2	&	19.00	&	18.93	&	18.88--19.18	&	0.16	&	0.15	&	0.12--0.20	&	0.0040	&	0.0040	&	 0.0010--0.0080	\\		
S3	&	19.00	&	18.98	&	18.98--19.18	&	0.16	&	0.15	&	0.11--0.19	&	0.0040	&	0.0040	&	 0.0040--0.0080	\\		
S4	&	18.50	&	18.48	&	18.48--18.68	&	0.18	&	0.18	&	0.12--0.24	&	0.0200	&	0.0200	&	 0.0100--0.0300	\\		
S5	&	18.30	&	18.30	&	18.28--18.46	&	0.14	&	0.14	&	0.08--0.20	&	0.0100	&	0.0100	&	 0.0080--0.0200	\\		
S6	&	19.20	&	19.24	&	19.18--19.37	&	0.13	&	0.12	&	0.09--0.17	&	0.0040	&	0.0040	&	 0.0010--0.0080	\\	
S7  &   18.40   &   18.40   &   18.30--18.50    &   0.08    &   0.08    &   0.00--0.18  &   0.0080  &   0.0080  &    0.0080 \\
S8  &   19.50   &   19.44   &   19.40--19.60    &   0.15    &   0.14    &   0.10--0.20  &   0.0080  &   0.0080  &    0.0080 \\ 	

 \hline
 \end{tabular}
 \end{table*}

 \begin{table*} 
 \caption{Similar to Table 2, but for other parameters. ``Age'' means the age of youngest stellar component.
 ``$Fb$'', ``$Fr$'', ``Nsf'' and ``mod'' mean binary fraction, rotating star fraction, number of star formation with interval of 0.1\,Gyr,
 and star formation mode (1, 2 and 3 corresponds to homogeneous, linearly increasing and linearly decreasing modes with increasing age).}
 \label{symbols}
 \begin{tabular}{cccccccccccccc}
  \hline\hline
No.	&Age$_{\rm in}$	&Age$_{\rm fit}$ &range	  &Fb$_{\rm in}$ &Fb$_{\rm fit}$ &range     &$Nsf_{\rm in}$	&$Nsf_{\rm fit}$	 &range	&mod$_{\rm in}$	&mod$_{\rm fit}$ &Fr$_{\rm in}$ &Fr$_{\rm fit}$ \\
        &[Gyr]       	&[Gyr]       	 &[Gyr]	  &            	 &            	 &          &              	&                 	 &     	&            	&	         &&\\
 \hline
S1	&	1.5	&	1.5	&	1.2--1.8	&	0.5	&	0.4	&	0.2--0.8	&	1	&	1	&	1--3	&	1	&	1	&0&0\\		
S2	&	0.5	&	0.5	&	0.2--0.8	&	0.7	&	0.8	&	0.4--1.0	&	2	&	3	&	1--3	&	1	&	2	&0&0\\		
S3	&	1.3	&	1.3	&	0.9--1.8	&	0.0	&	0.2	&	0.0--0.3	&	2	&	3	&	1--3	&	1	&	2	&0&0\\		
S4	&	0.6	&	0.6	&	0.3--0.9	&	0.5	&	0.7	&	0.2--0.8	&	3	&	3	&	1--3	&	1	&	1	&0&0\\		
S5	&	1.0	&	1.0	&	0.8--1.1	&	0.5	&	0.4	&	0.3--0.8	&	1	&	1	&	1--1	&	1	&	1	&0&0\\		
S6	&	0.8	&	0.9	&	0.5--1.1	&	0.0	&	0.0	&	0.0--0.0	&	4	&	3	&	2--5	&	1	&	3	&0&0\\		
S7  &   1.0 &   1.0 &   0.5--1.5    &   0.5 &   0.4 &   0.3--0.7    &   1   &   1   &   1--1    &   1   &   1   & 1.0 & 1.0 \\
S8  &   0.8 &   0.8 &   0.5--1.1    &   0.3 &   0.1 &   0.1--0.5    &   1   &   1   &   1--1    &   1   &   1   & 0.3 & 0.3 \\
 \hline
 \end{tabular}
 \end{table*}

 \begin{figure}
\begin{minipage}[t]{0.5\linewidth}
\centering
\includegraphics[width=\textwidth]{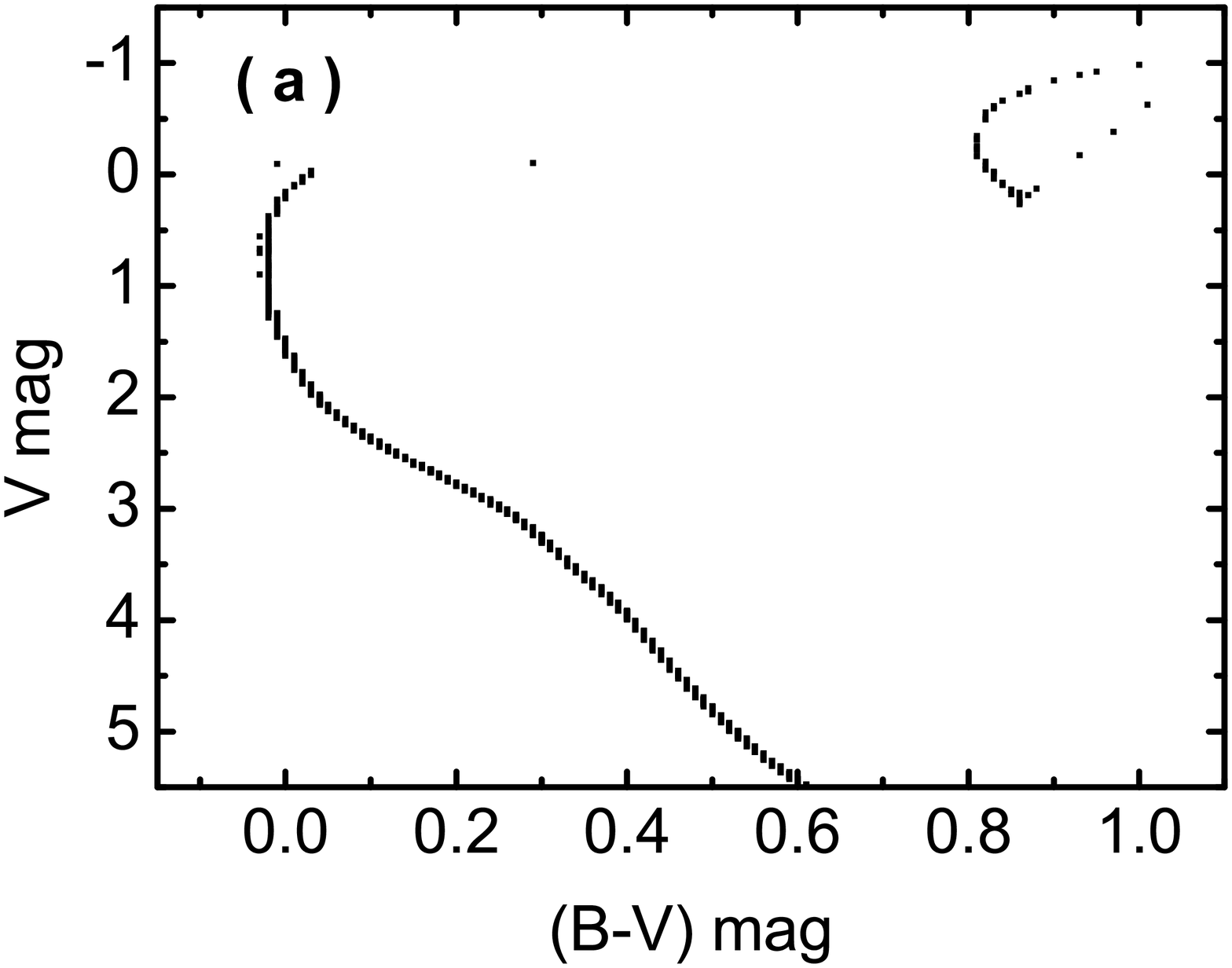}
\end{minipage}%
\begin{minipage}[t]{0.5\linewidth}
\centering
\includegraphics[width=\textwidth]{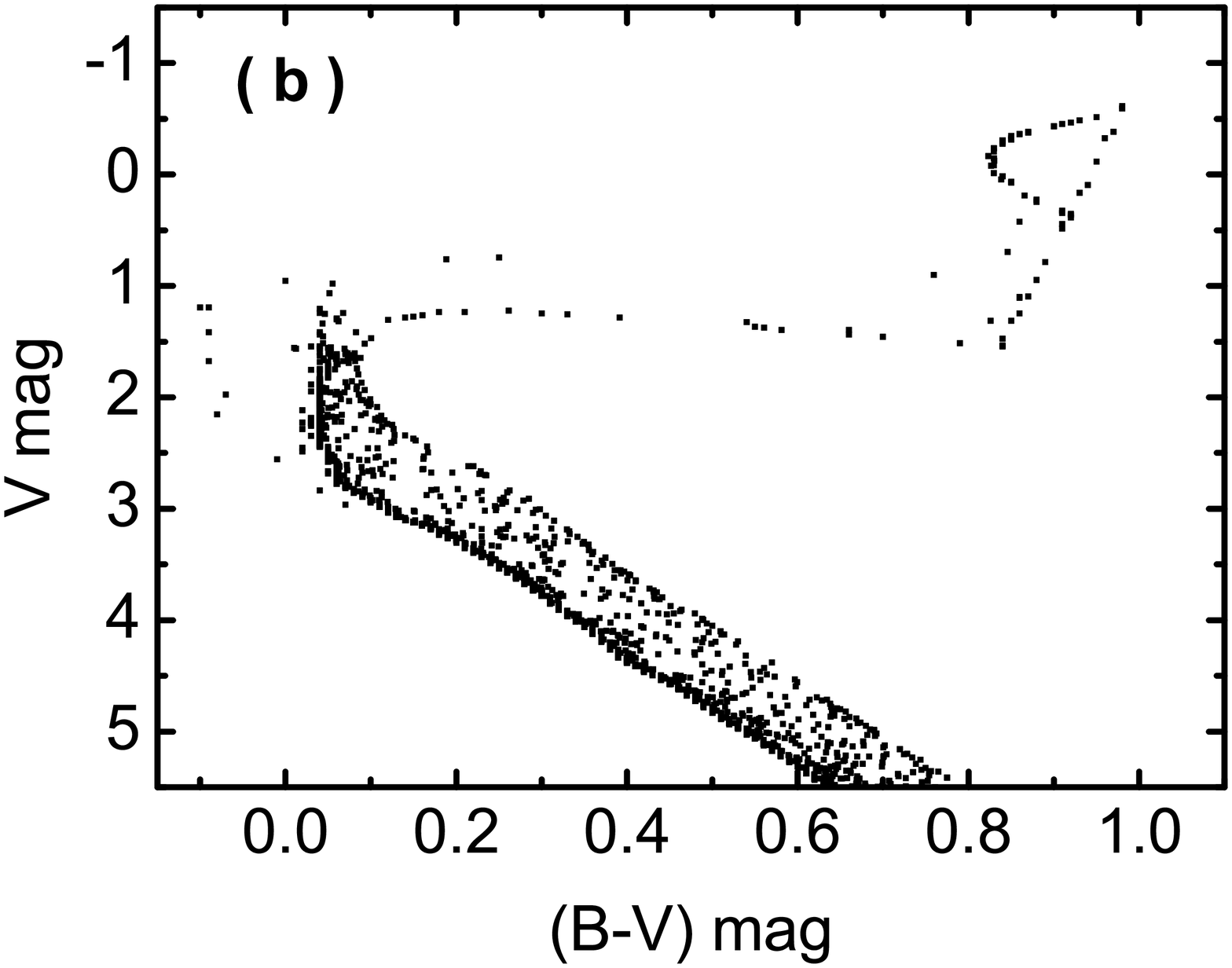}
\end{minipage}
\begin{minipage}[t]{0.5\linewidth}
\centering
\includegraphics[width=\textwidth]{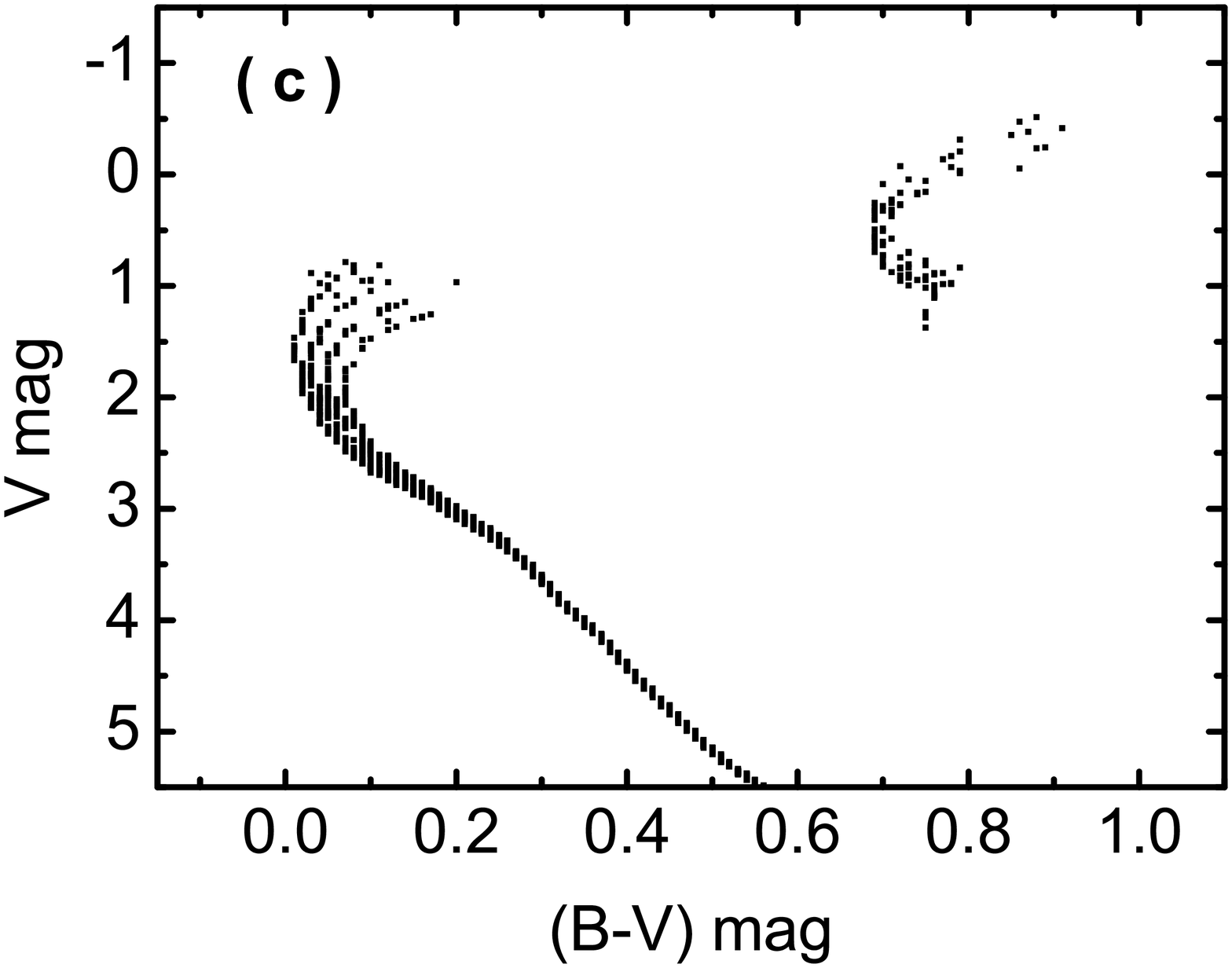}
\end{minipage}%
\begin{minipage}[t]{0.5\linewidth}
\centering
\includegraphics[width=\textwidth]{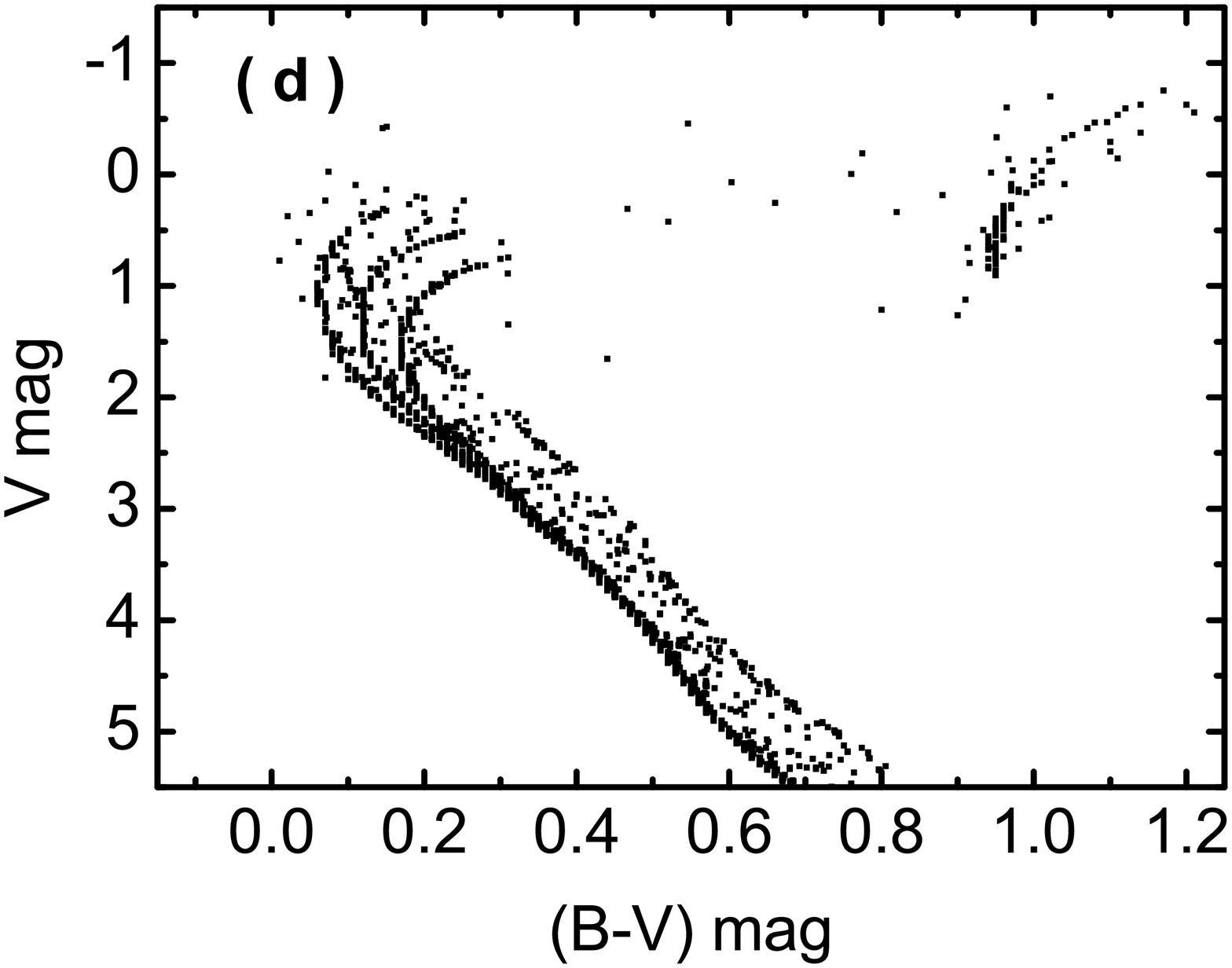}
\end{minipage}
\begin{minipage}[t]{0.5\linewidth}
\centering
\includegraphics[width=\textwidth]{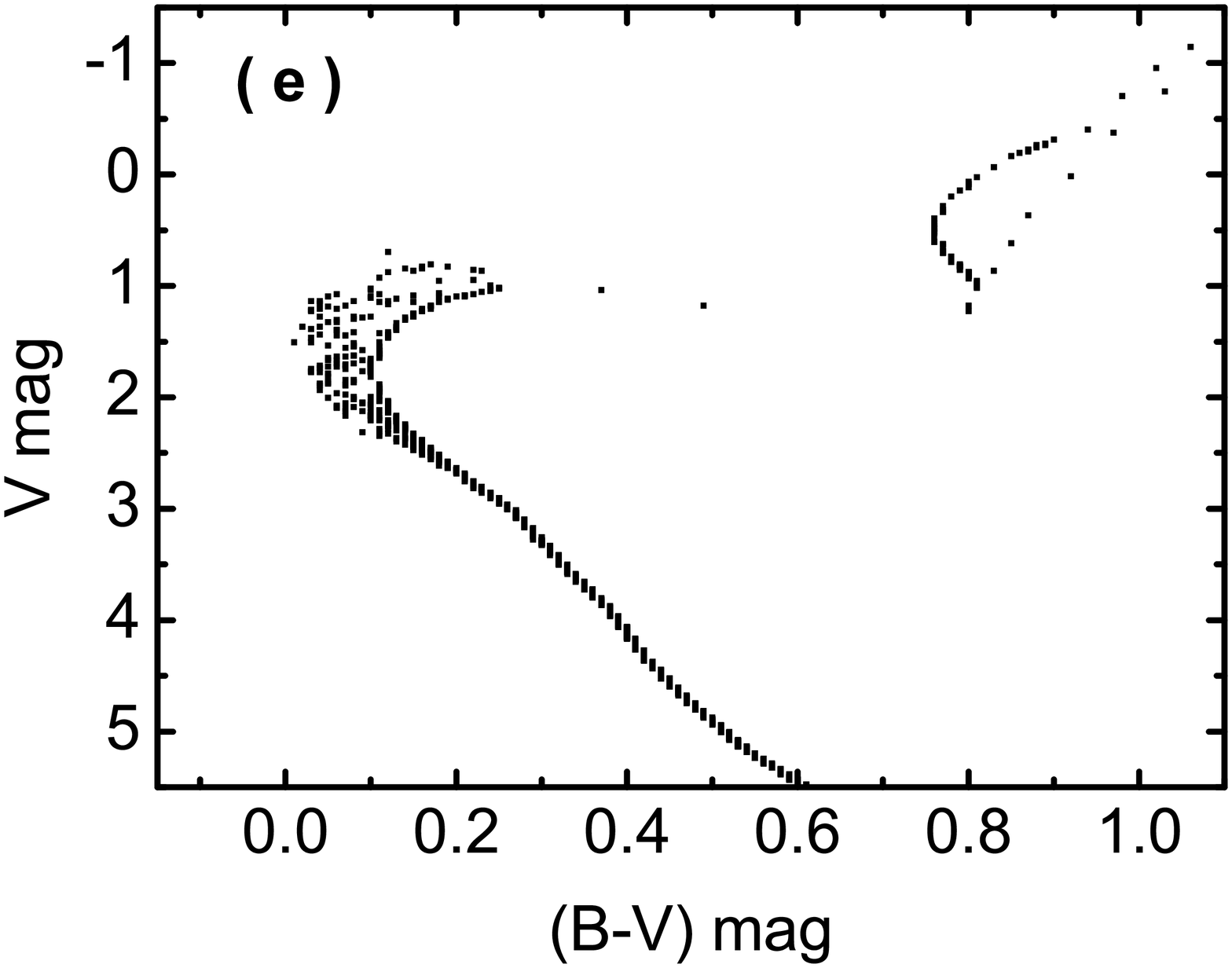}
\end{minipage}%
\begin{minipage}[t]{0.5\linewidth}
\centering
\includegraphics[width=\textwidth]{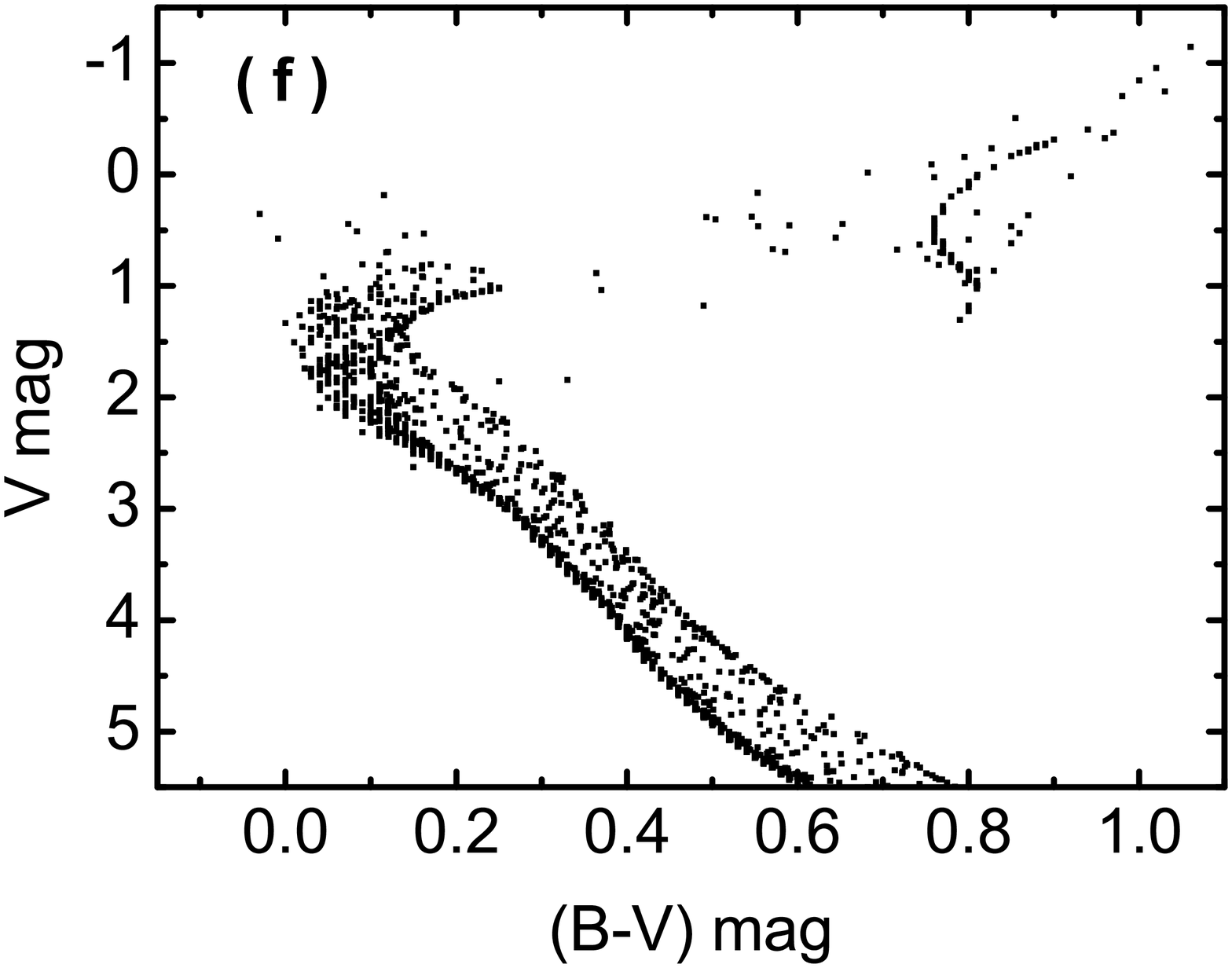}
\end{minipage}
\begin{minipage}[t]{0.5\linewidth}
\centering
\includegraphics[width=\textwidth]{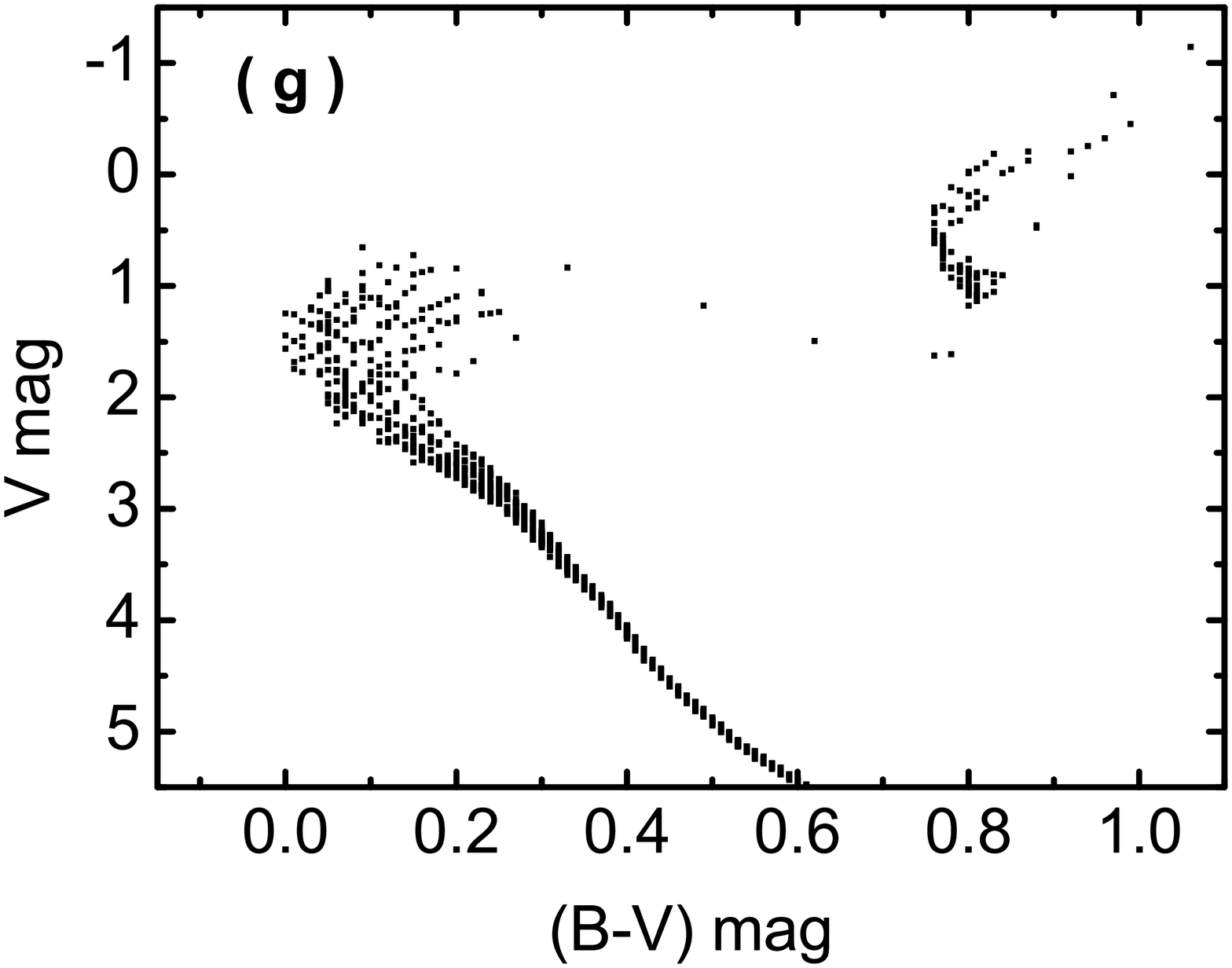}
\end{minipage}%
\begin{minipage}[t]{0.5\linewidth}
\centering
\includegraphics[width=\textwidth]{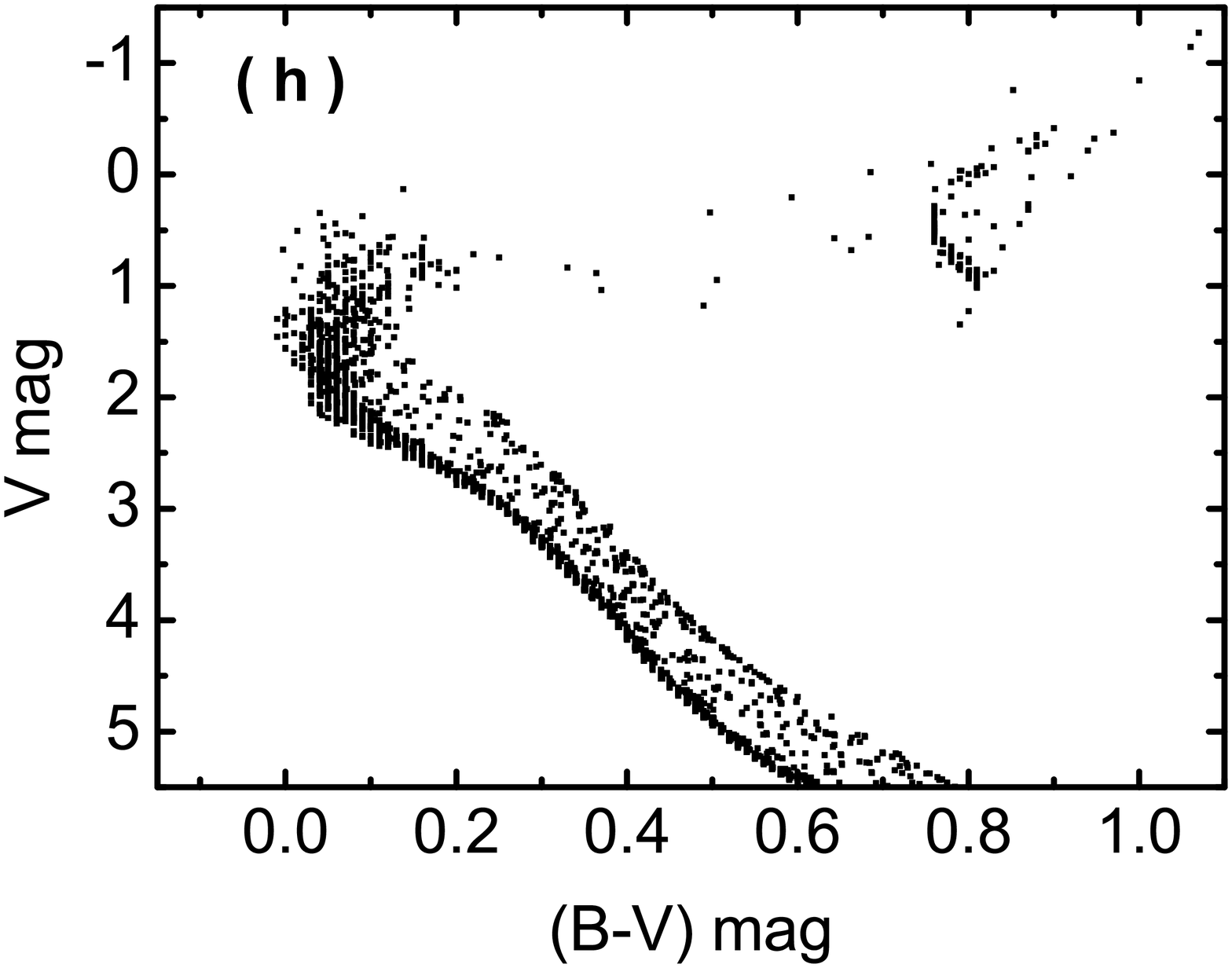}
\end{minipage}
  \caption{Examples for intrinsic CMDs of various stellar populations. All CMDs are built via \emph{Powerful CMD}. Model inputs are in Table 1.}
\end{figure}

\begin{figure}
\begin{minipage}[t]{\linewidth}
\centering
\includegraphics[width=\textwidth]{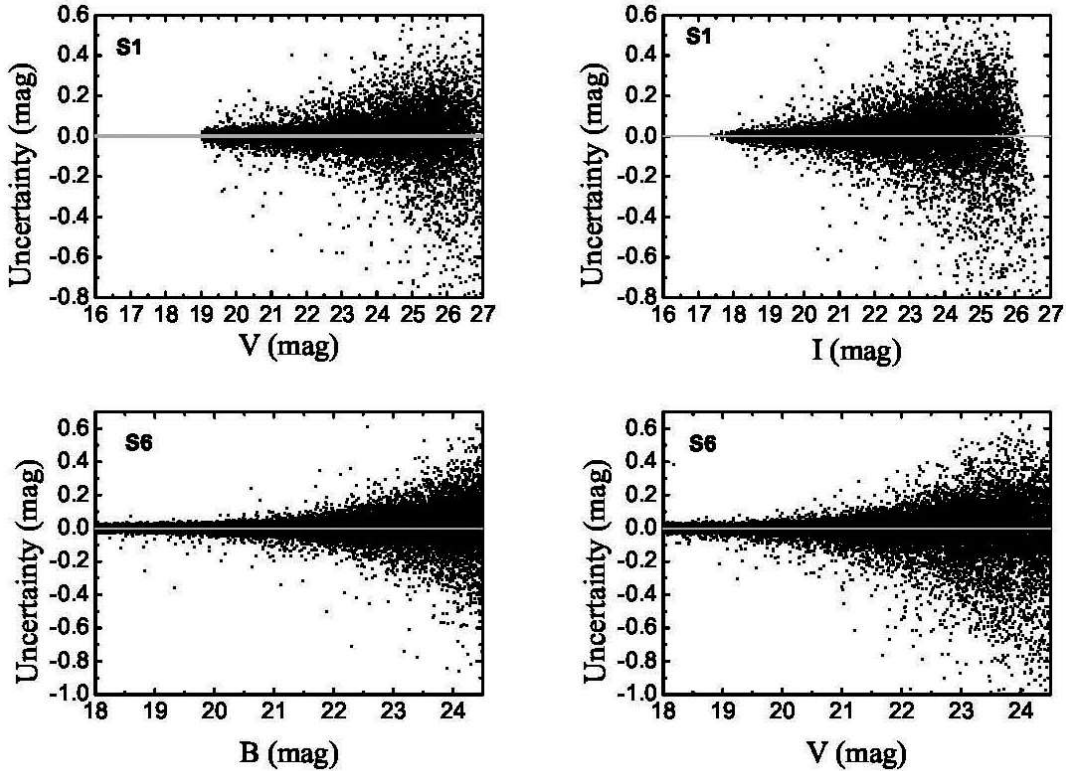}
\end{minipage}
  \caption{Magnitude uncertainty as a function of observed magnitude of two simulated star clusters.
  The uncertainties are estimated via ASTs,
  and the errors relating to observational equipment are not taken into account because they are usually much smaller.
  }
\end{figure}

\begin{figure}
\begin{minipage}[t]{0.5\linewidth}
\centering
\includegraphics[width=\textwidth]{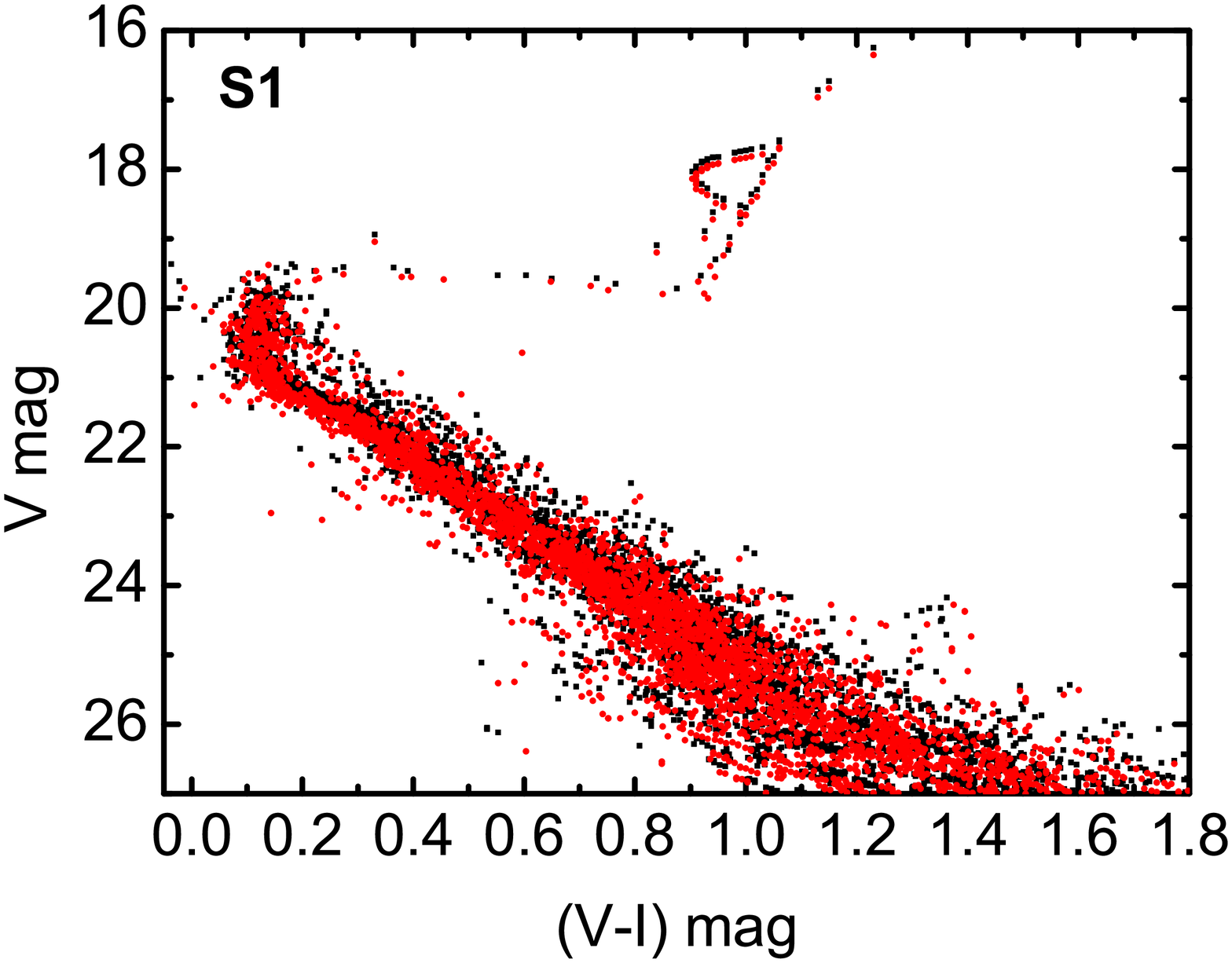}
\end{minipage}%
\begin{minipage}[t]{0.5\linewidth}
\centering
\includegraphics[width=\textwidth]{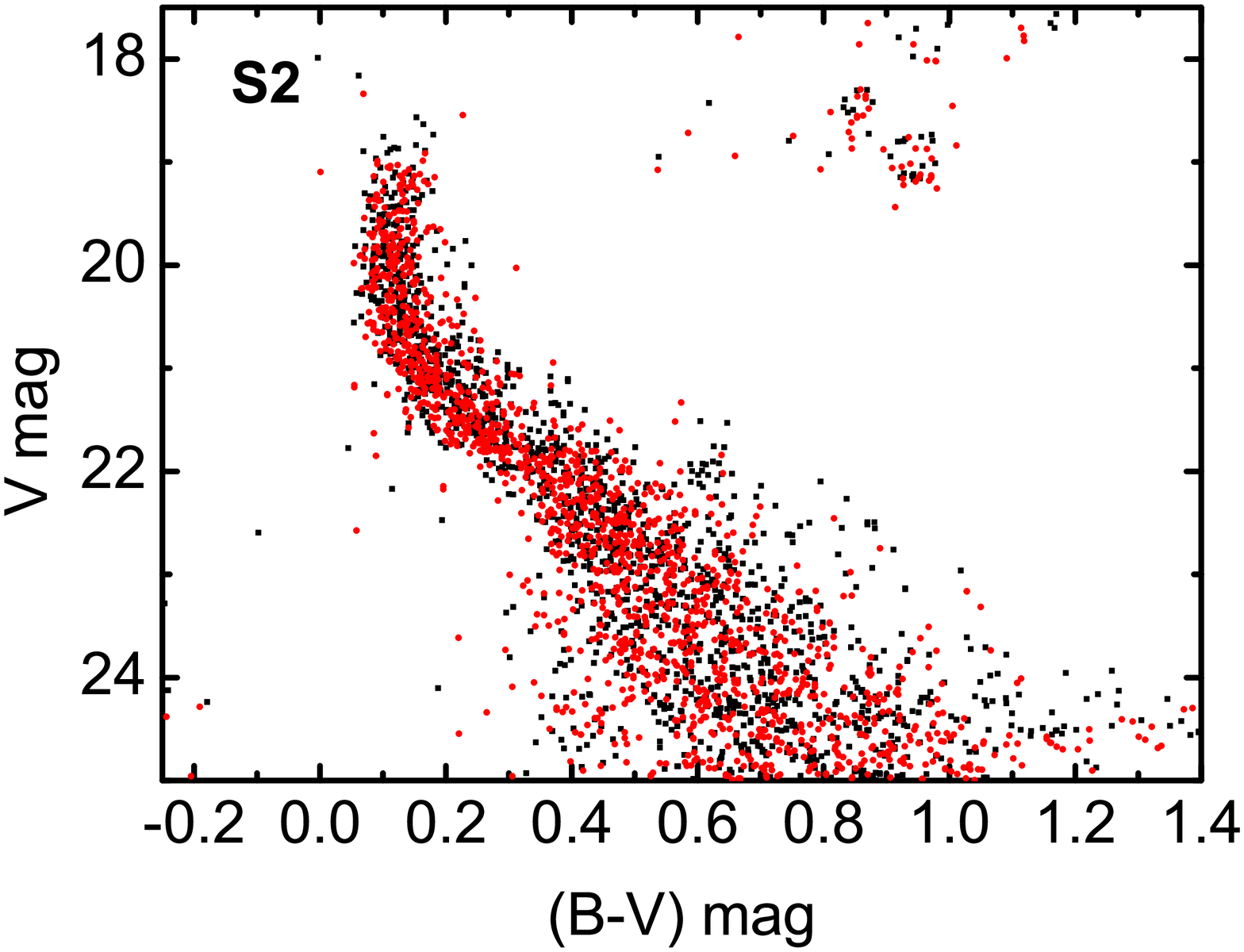}
\end{minipage}
\begin{minipage}[t]{0.5\linewidth}
\centering
\includegraphics[width=\textwidth]{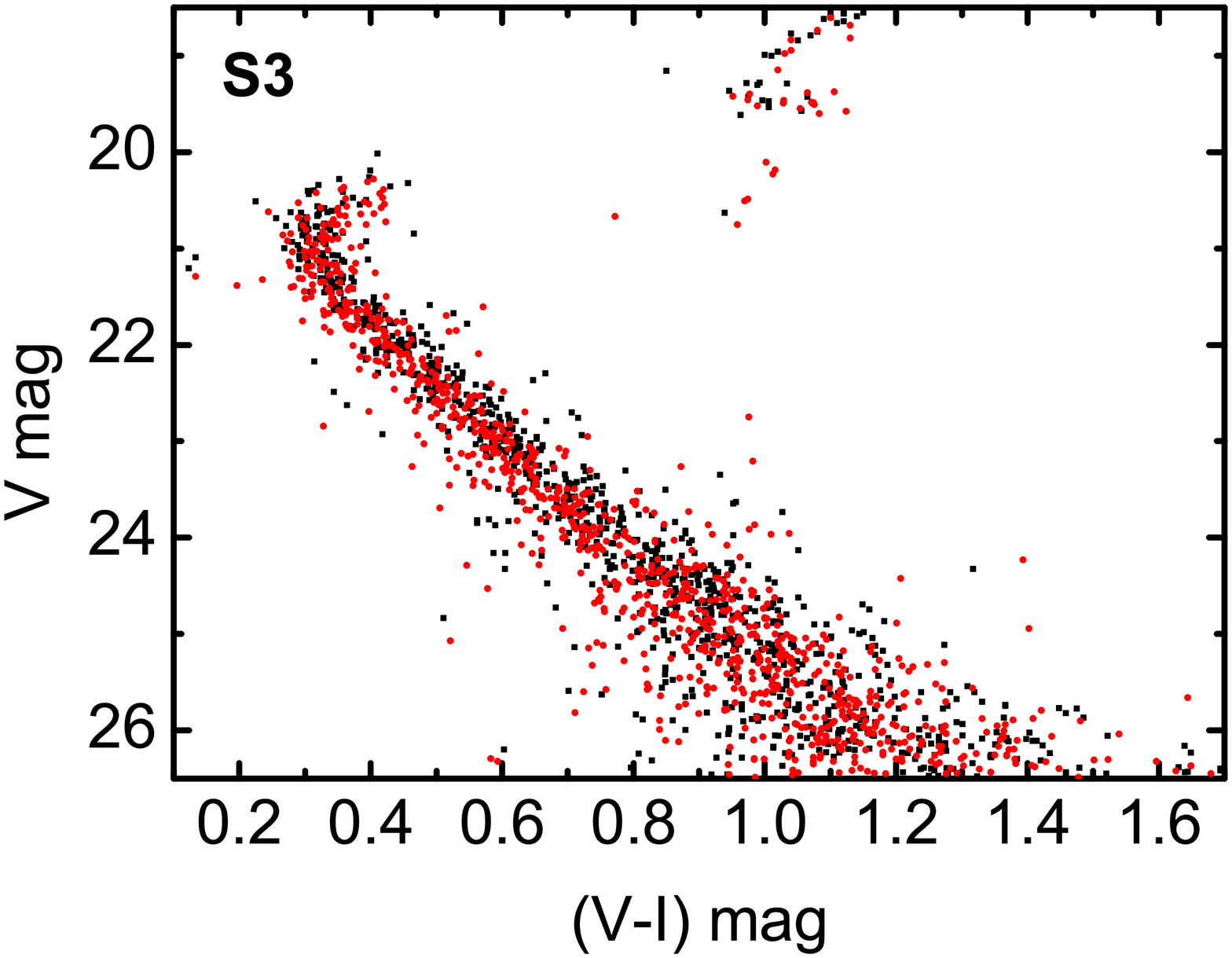}
\end{minipage}%
\begin{minipage}[t]{0.5\linewidth}
\centering
\includegraphics[width=\textwidth]{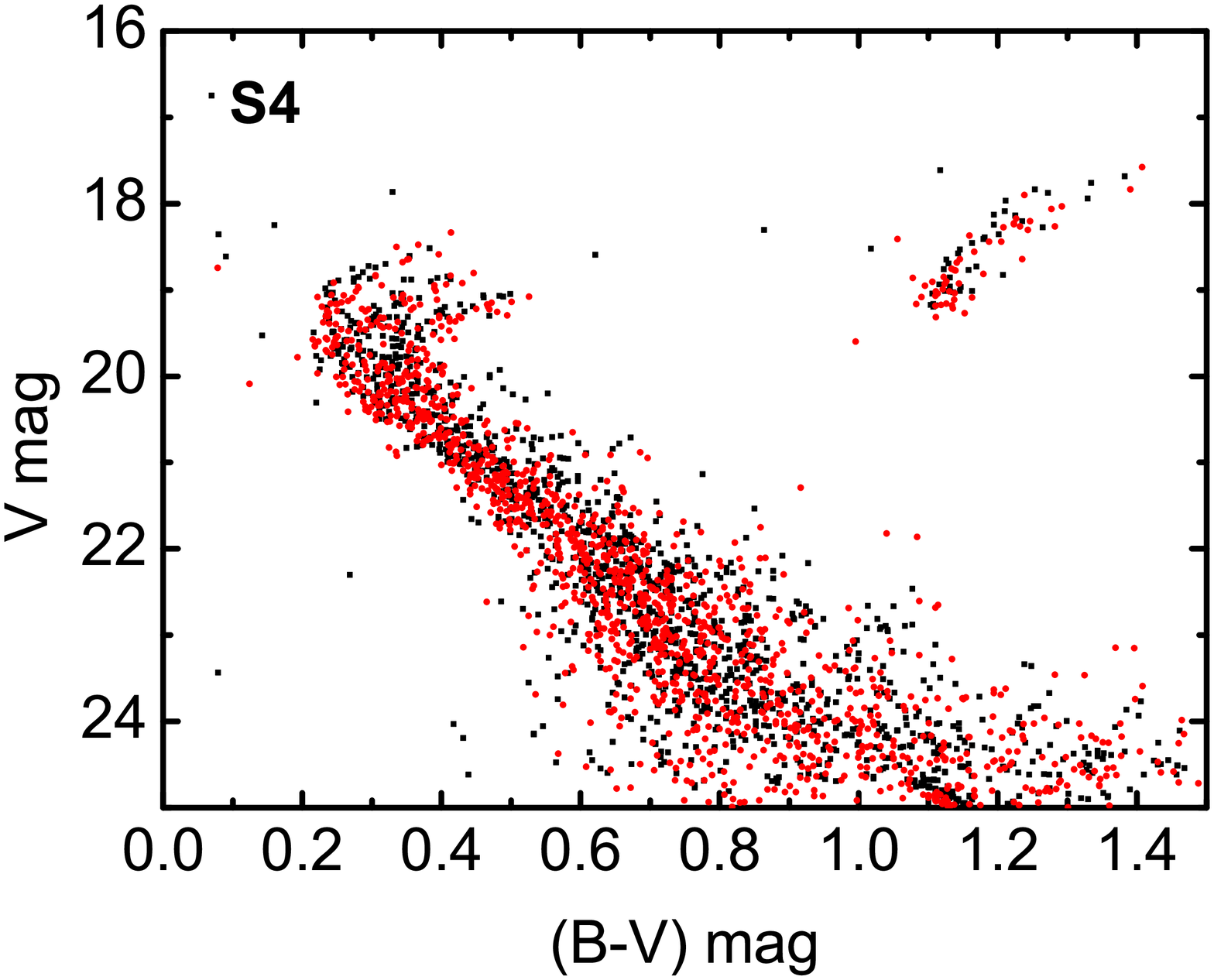}
\end{minipage}
\begin{minipage}[t]{0.5\linewidth}
\centering
\includegraphics[width=\textwidth]{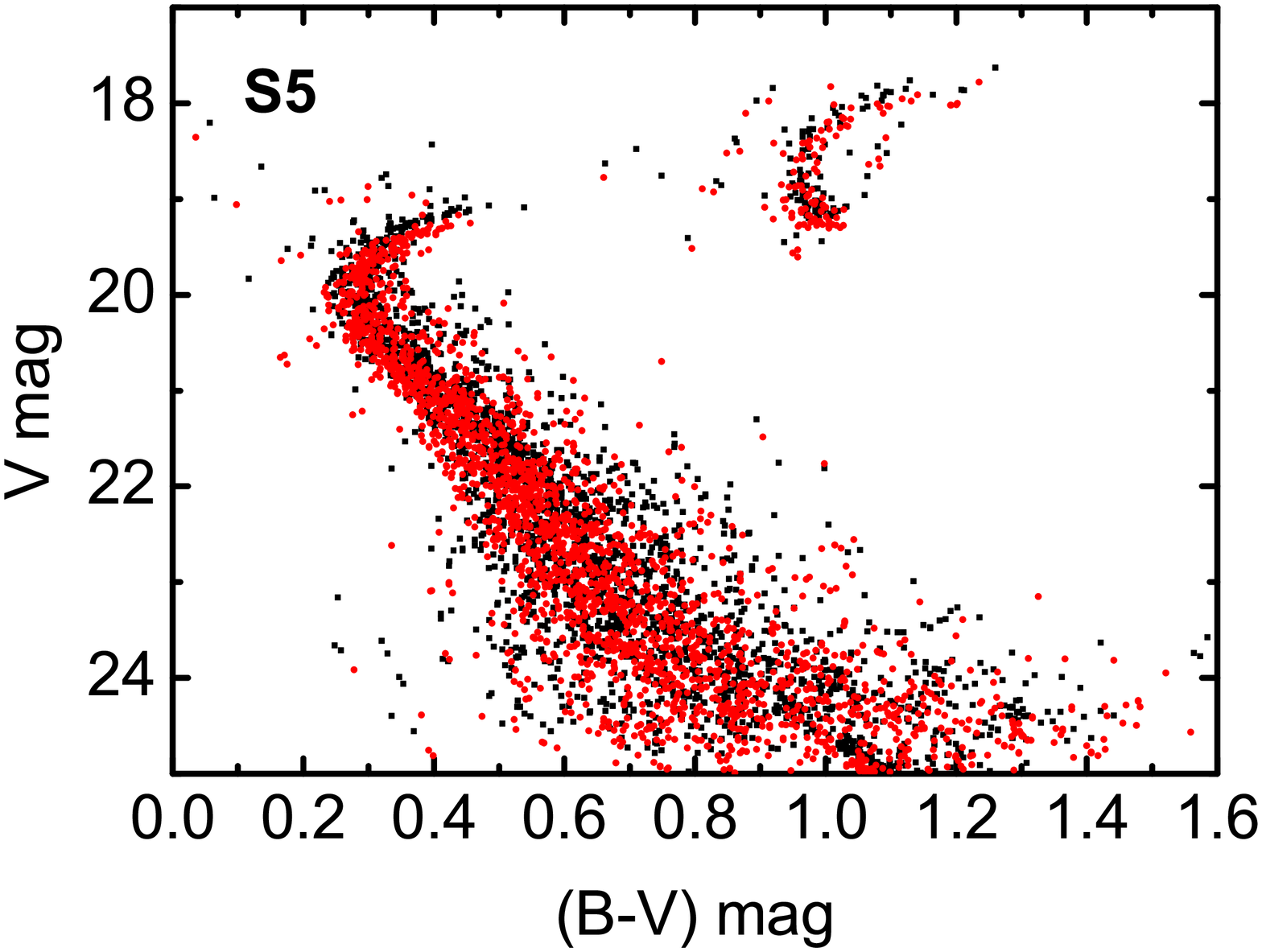}
\end{minipage}%
\begin{minipage}[t]{0.5\linewidth}
\centering
\includegraphics[width=\textwidth]{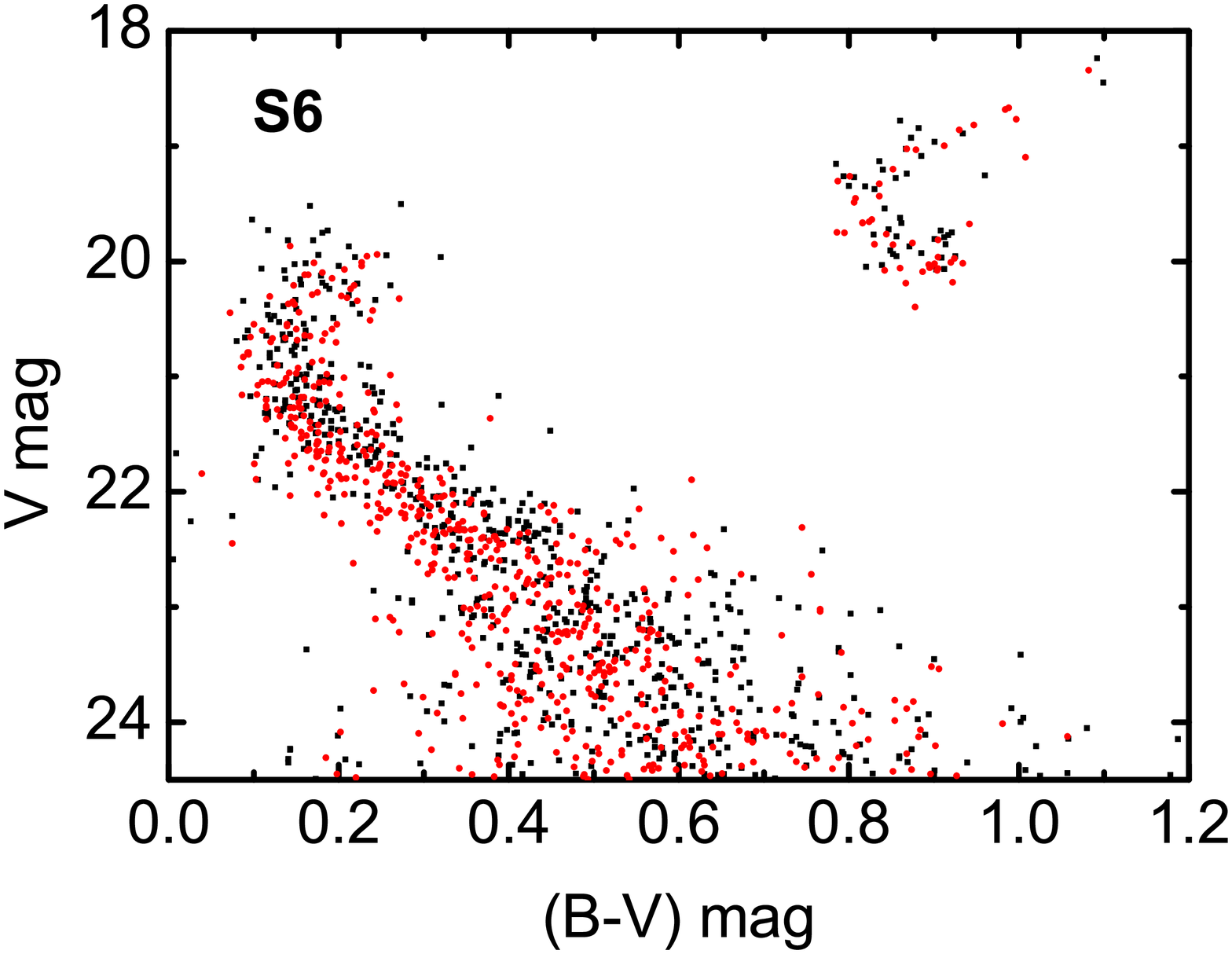}
\end{minipage}%

\begin{minipage}[t]{0.5\linewidth}
\centering
\includegraphics[width=\textwidth]{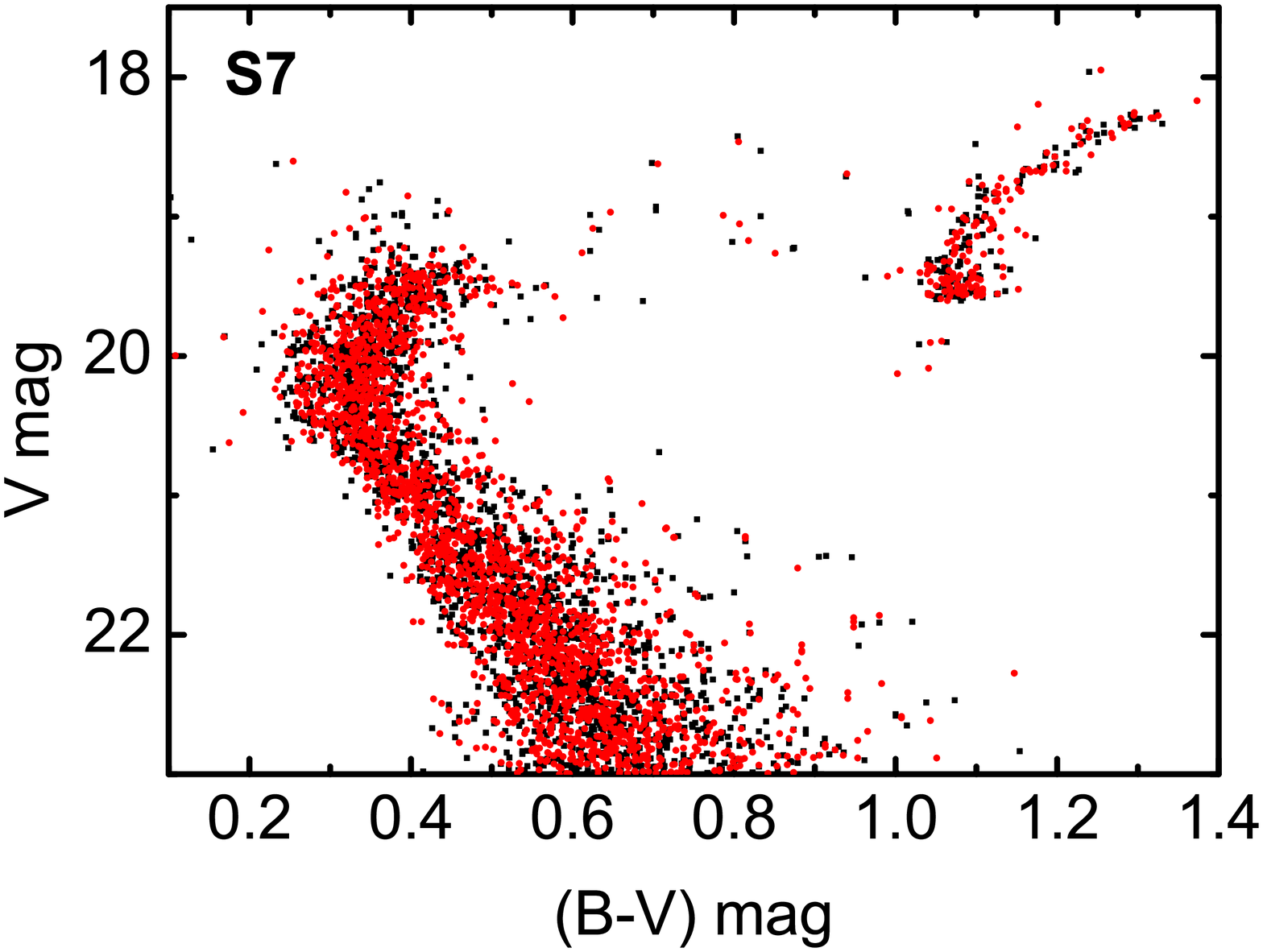}
\end{minipage}%
\begin{minipage}[t]{0.5\linewidth}
\centering
\includegraphics[width=\textwidth]{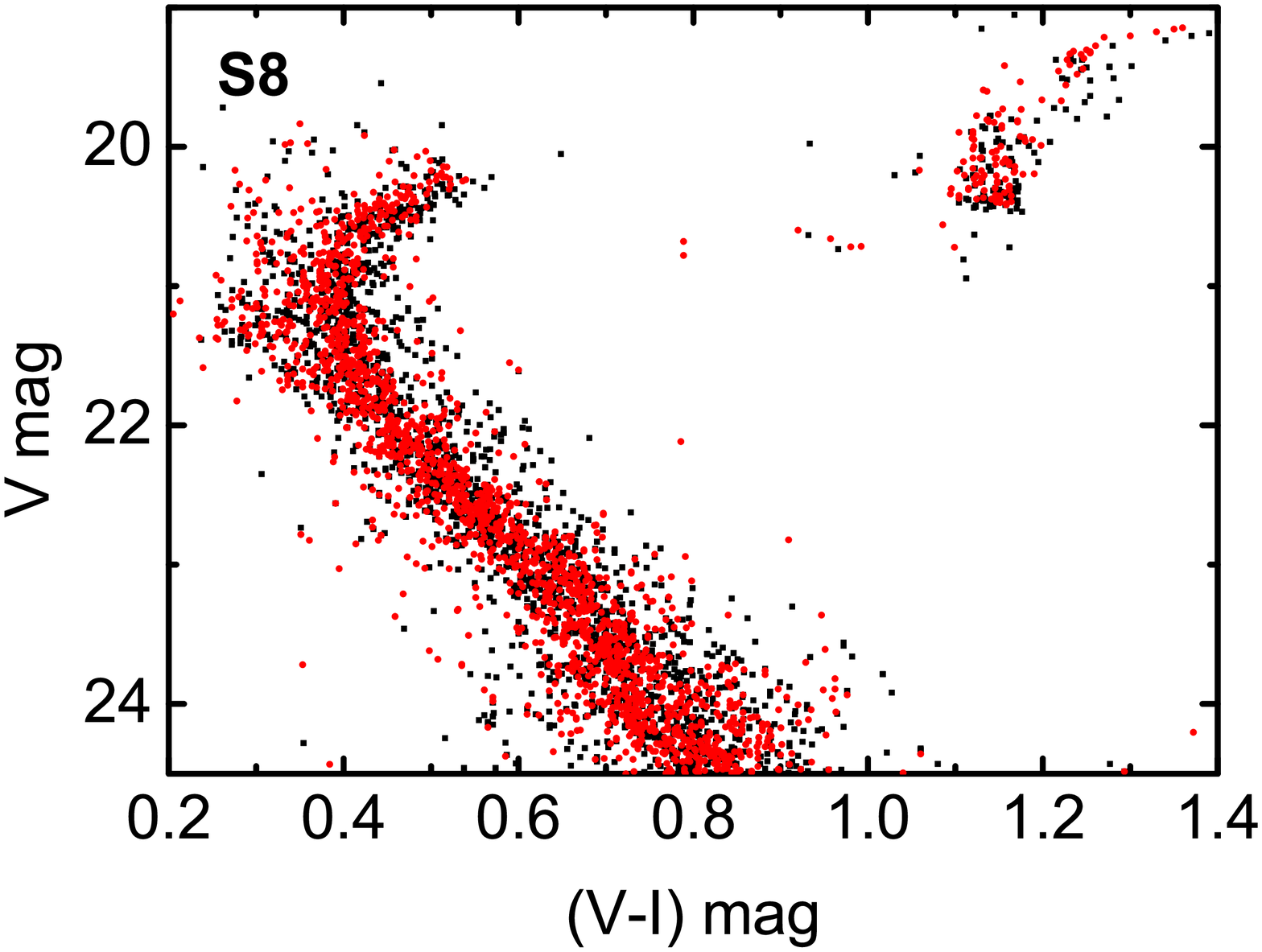}
\end{minipage}

\caption{Example CMDs (black) of simulated stellar populations and their best-fit CMDs (red).
  The best-fit CMDs are found by \emph{Powerful CMD}.}
\end{figure}

\begin{figure}
\begin{minipage}[t]{0.5\linewidth}
\centering
\includegraphics[width=\textwidth]{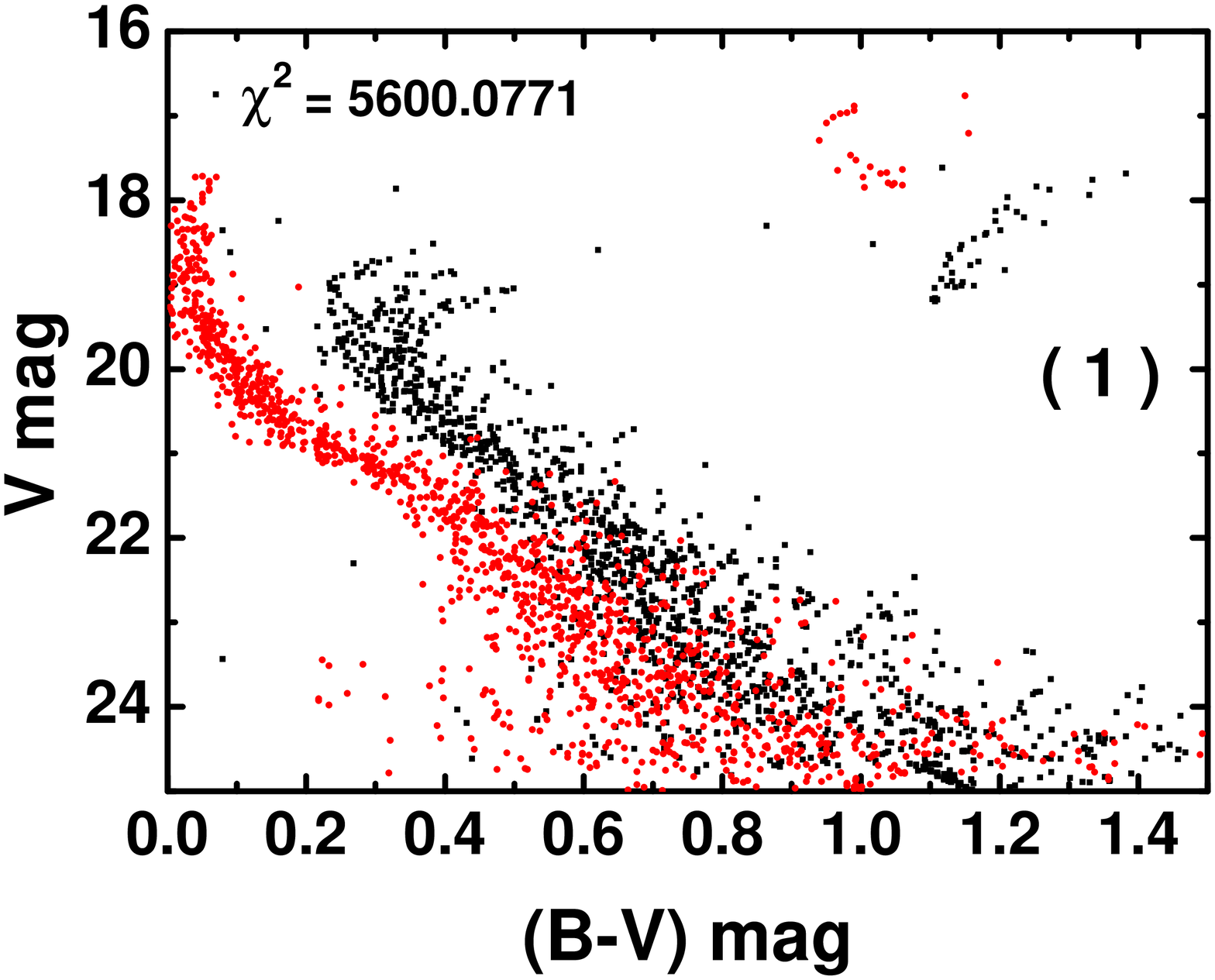}
\end{minipage}%
\begin{minipage}[t]{0.5\linewidth}
\centering
\includegraphics[width=\textwidth]{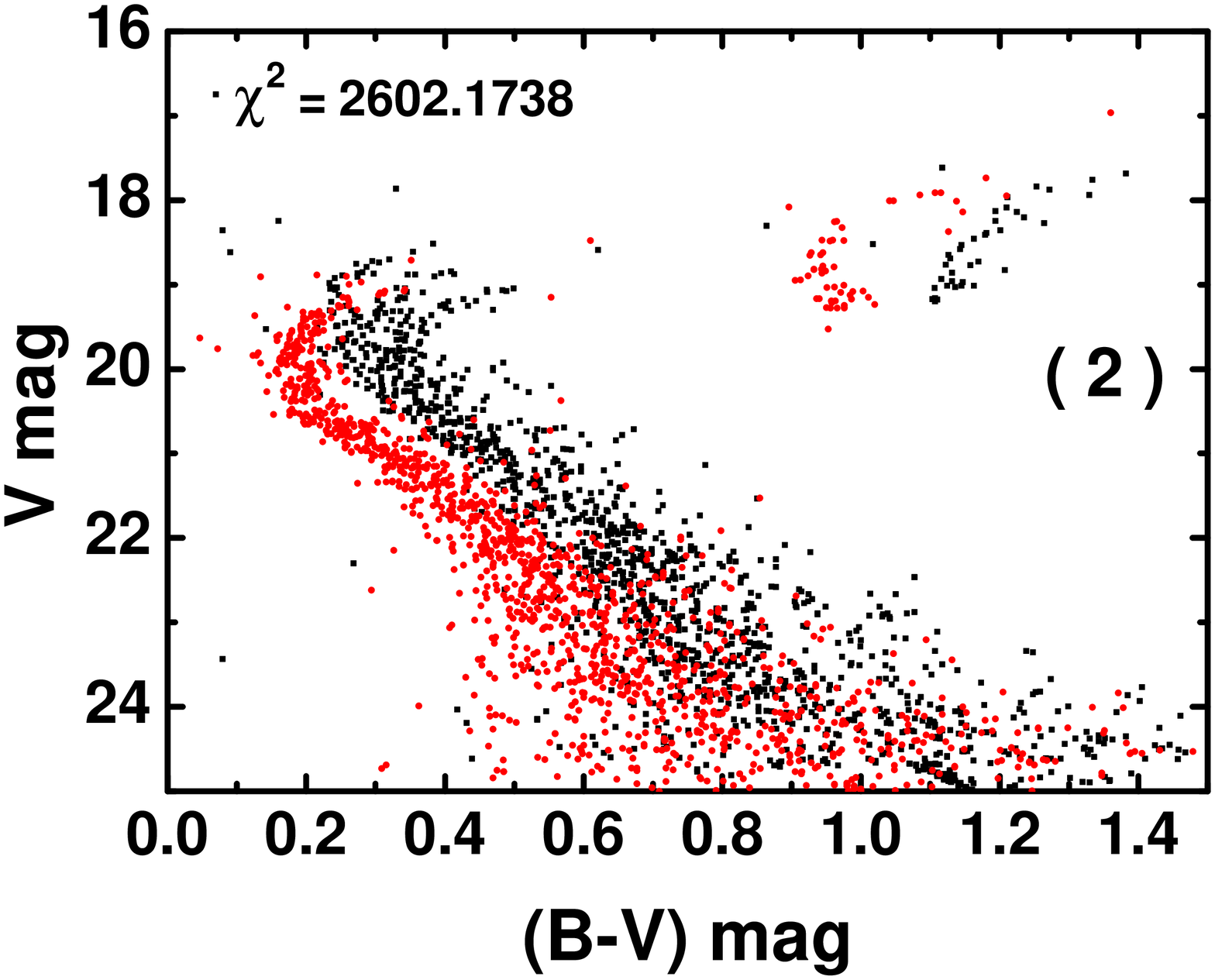}
\end{minipage}
\begin{minipage}[t]{0.5\linewidth}
\centering
\includegraphics[width=\textwidth]{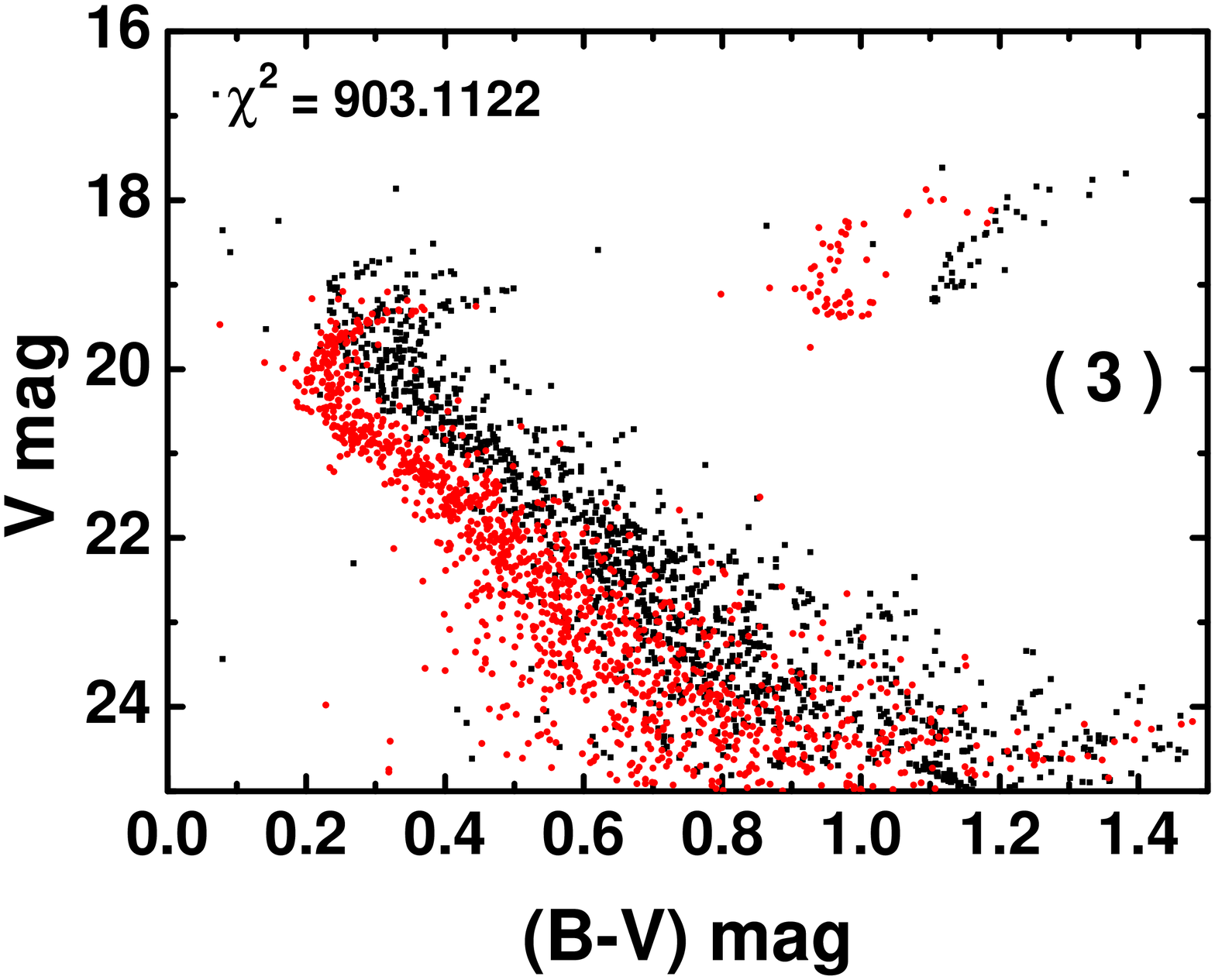}
\end{minipage}%
\begin{minipage}[t]{0.5\linewidth}
\centering
\includegraphics[width=\textwidth]{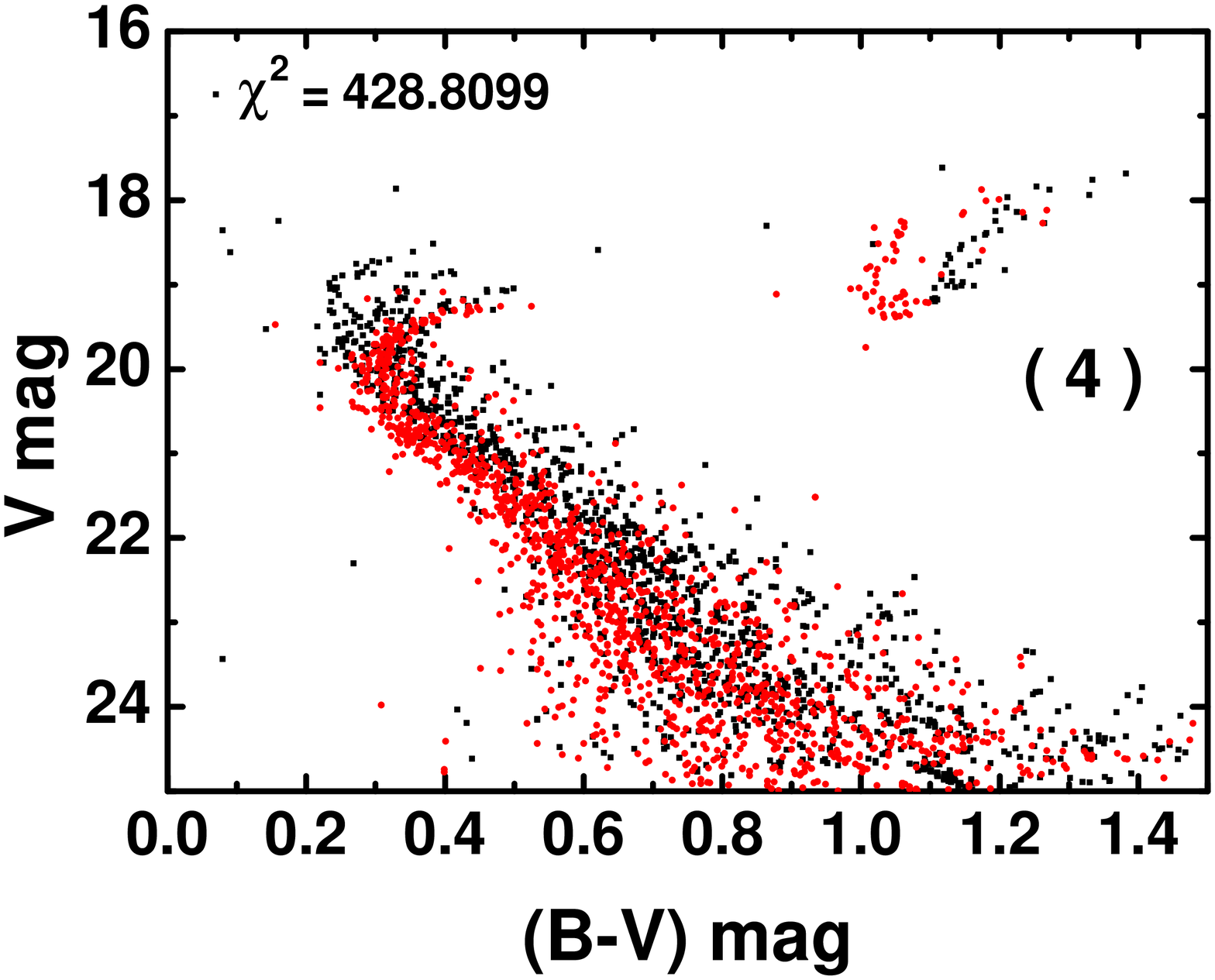}
\end{minipage}
\begin{minipage}[t]{0.5\linewidth}
\centering
\includegraphics[width=\textwidth]{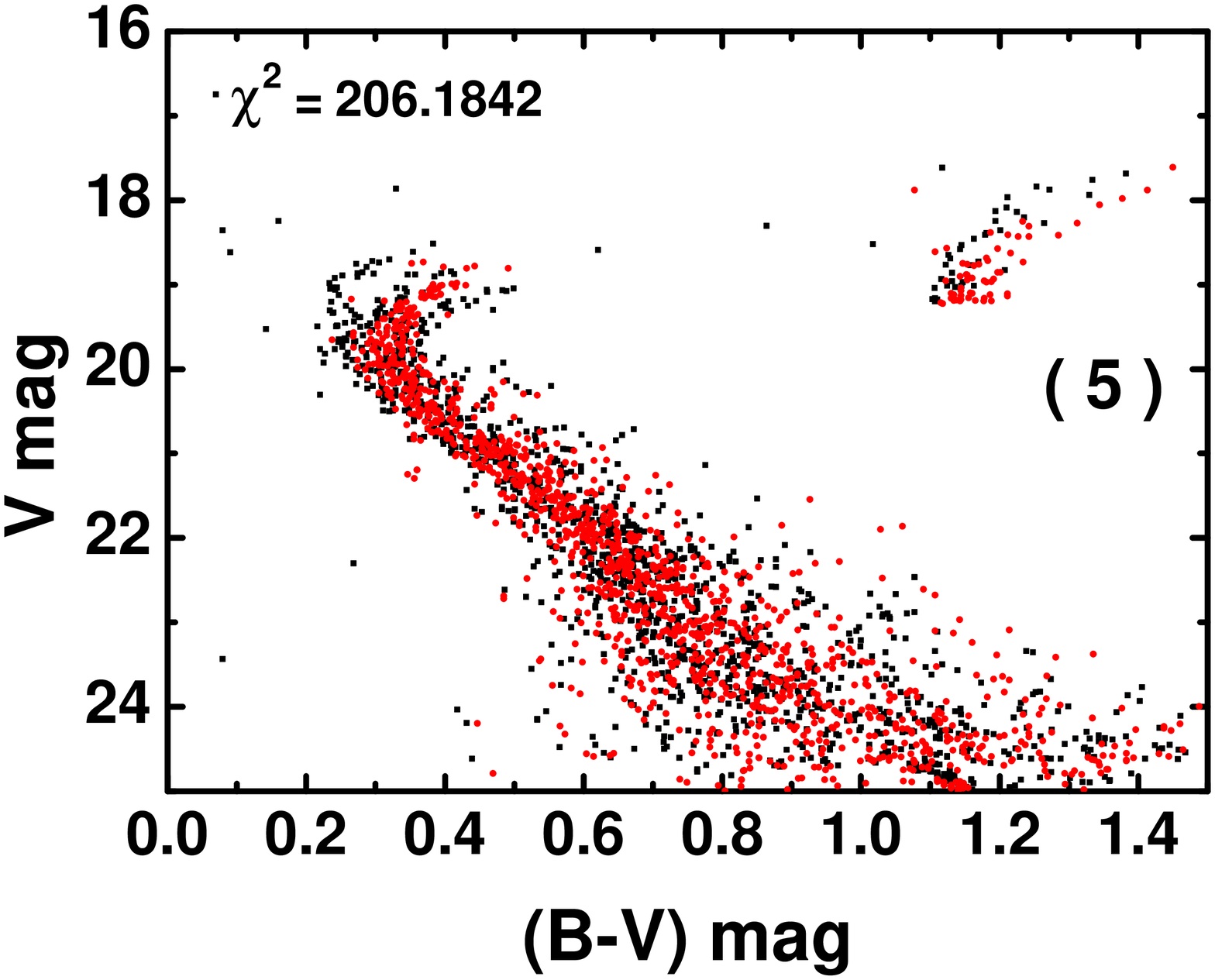}
\end{minipage}%
\begin{minipage}[t]{0.5\linewidth}
\centering
\includegraphics[width=\textwidth]{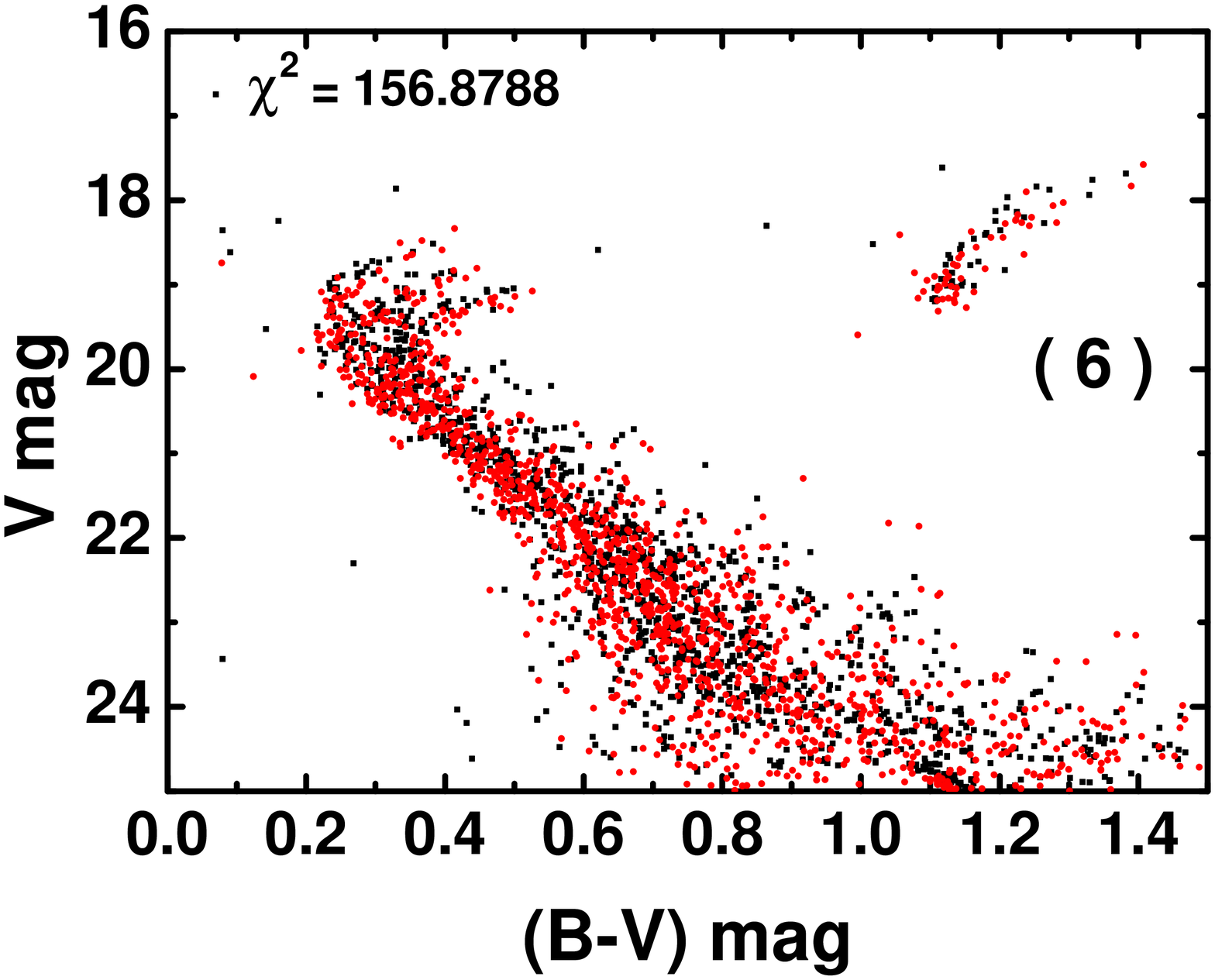}
\end{minipage}
  \caption{Process of \emph{Powerful CMD} to find the best-fit stellar population models.
  Black and red points are for observed and fitted CMDs, respectively.
  The less the $\chi^2$, the better the goodness of fit. $\chi^2$ less than 431 denotes the acceptable models.}
\end{figure}

\begin{figure}
\begin{minipage}[t]{0.5\linewidth}
\centering
\includegraphics[width=0.7\textwidth]{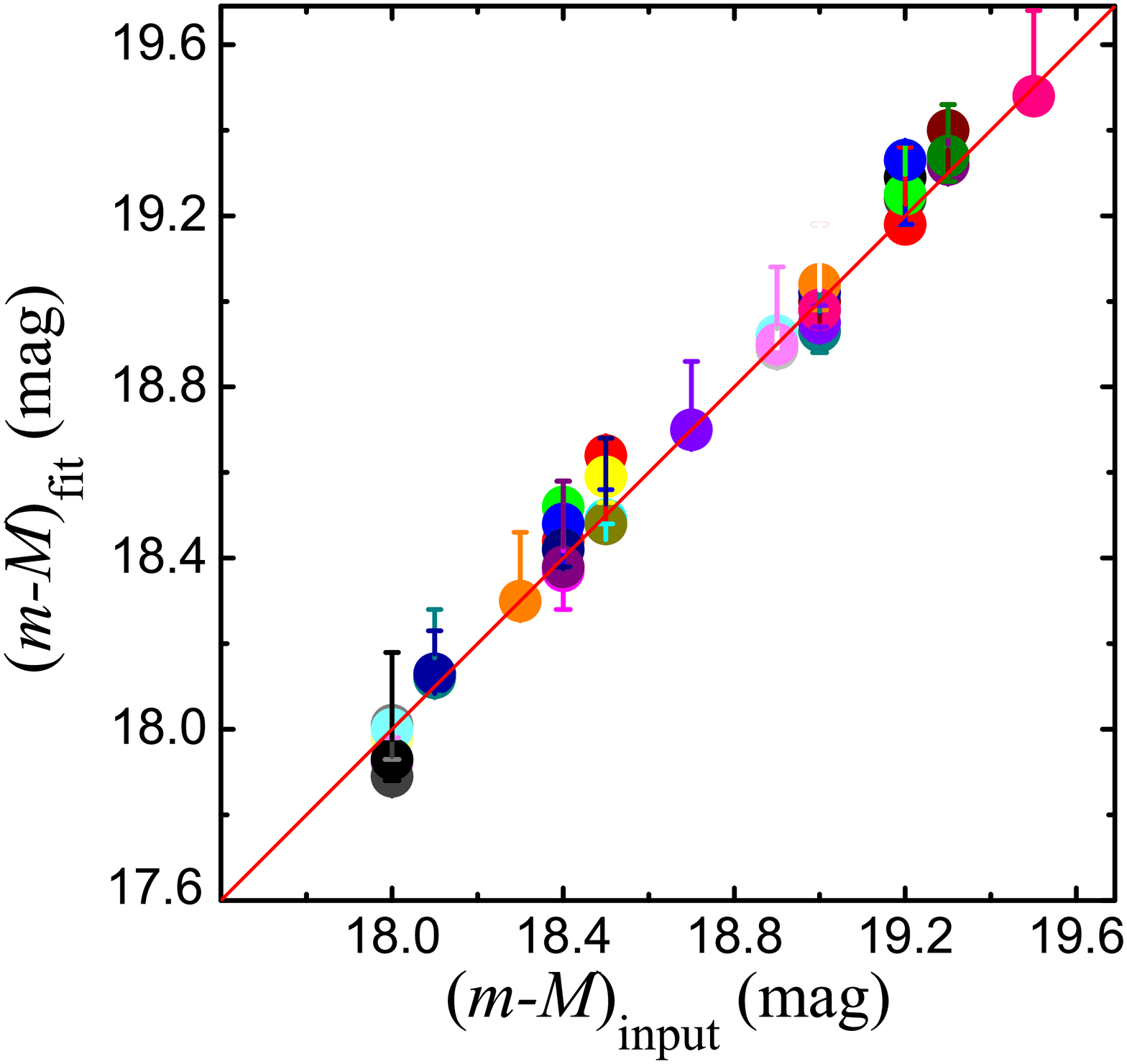}
\end{minipage}%
\begin{minipage}[t]{0.5\linewidth}
\centering
\includegraphics[width=0.7\textwidth]{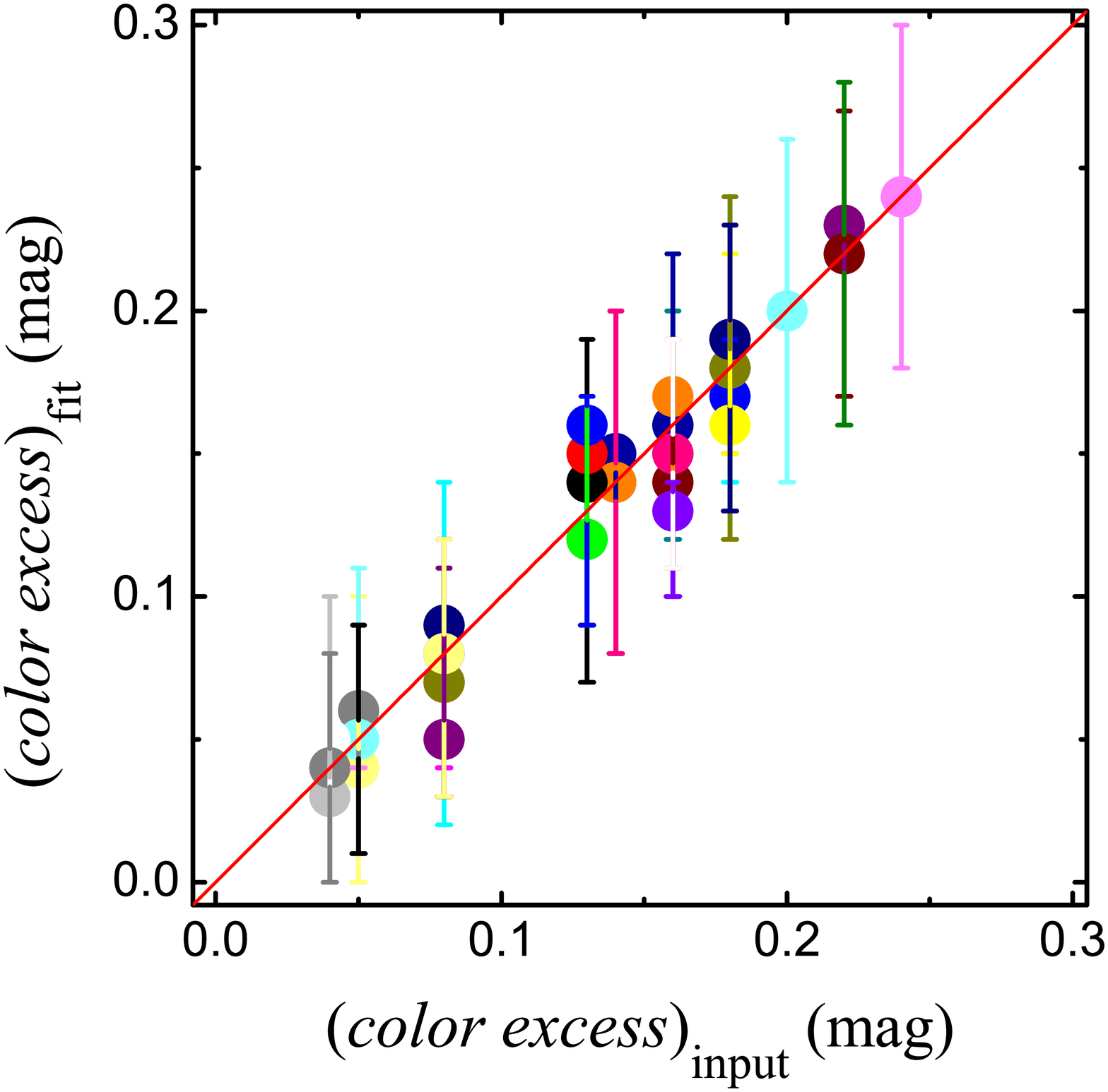}
\end{minipage}
\begin{minipage}[t]{0.5\linewidth}
\centering
\includegraphics[width=0.7\textwidth]{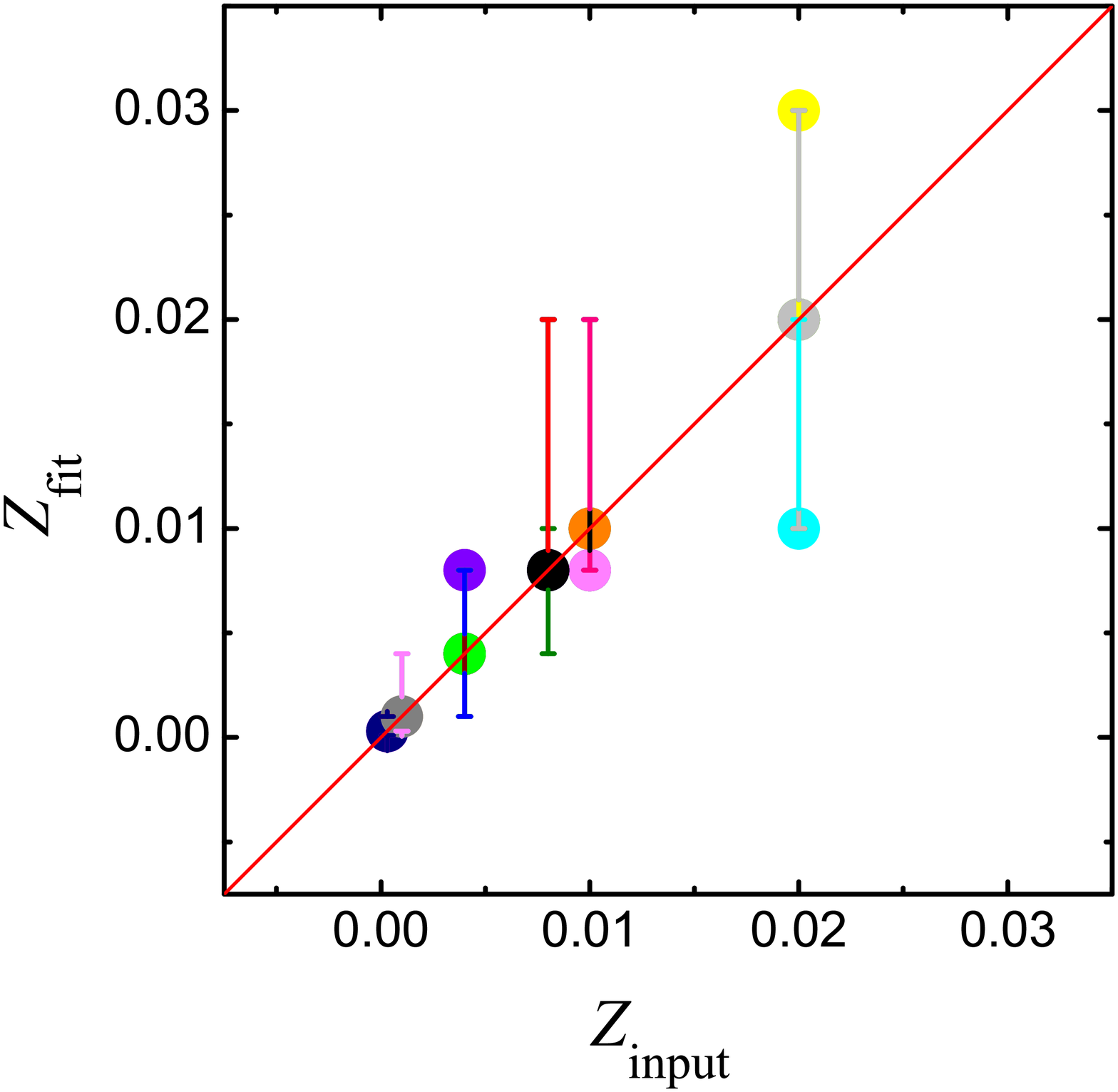}
\end{minipage}%
\begin{minipage}[t]{0.5\linewidth}
\centering
\includegraphics[width=0.7\textwidth]{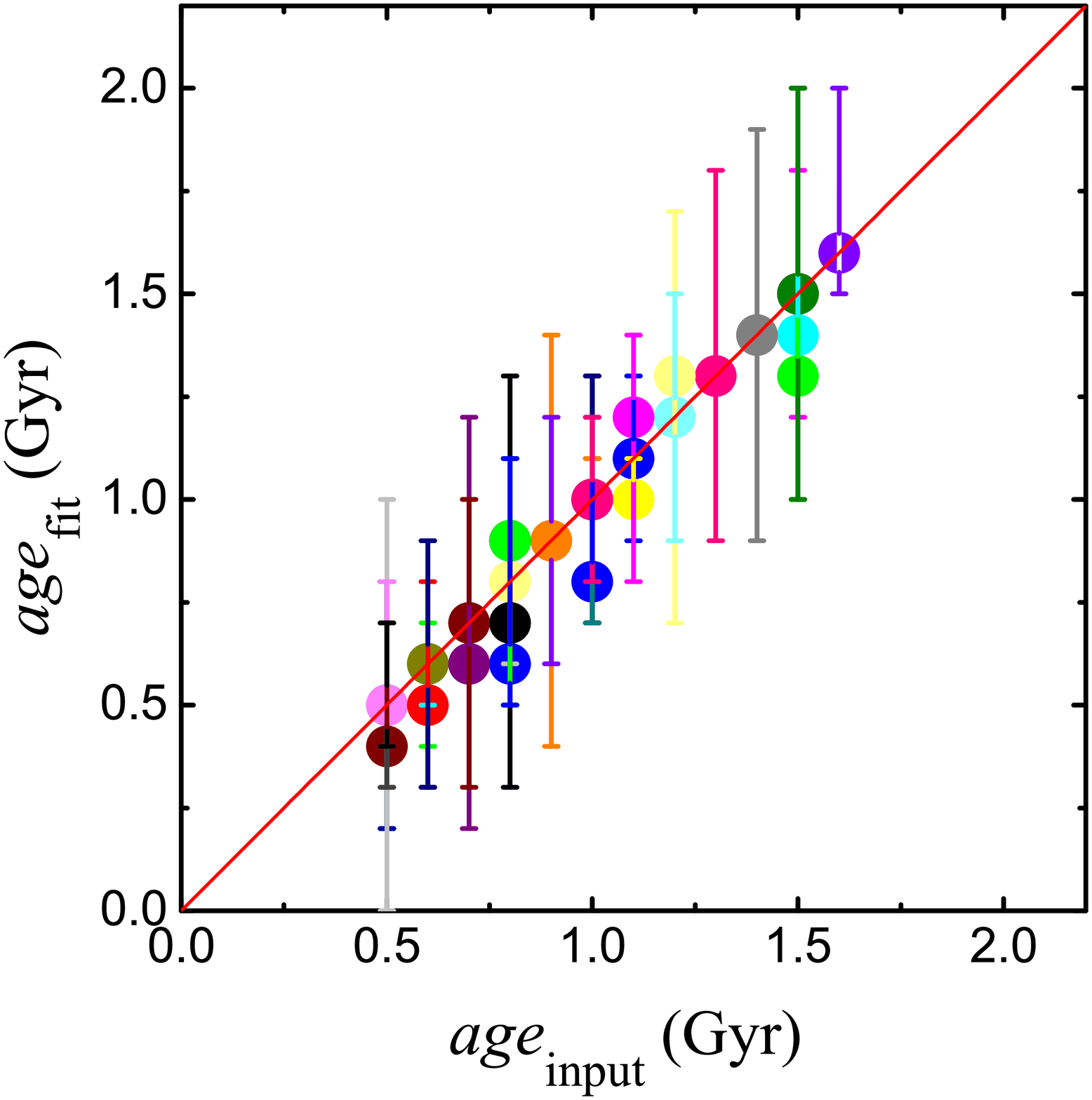}
\end{minipage}
\begin{minipage}[t]{0.5\linewidth}
\centering
\includegraphics[width=0.7\textwidth]{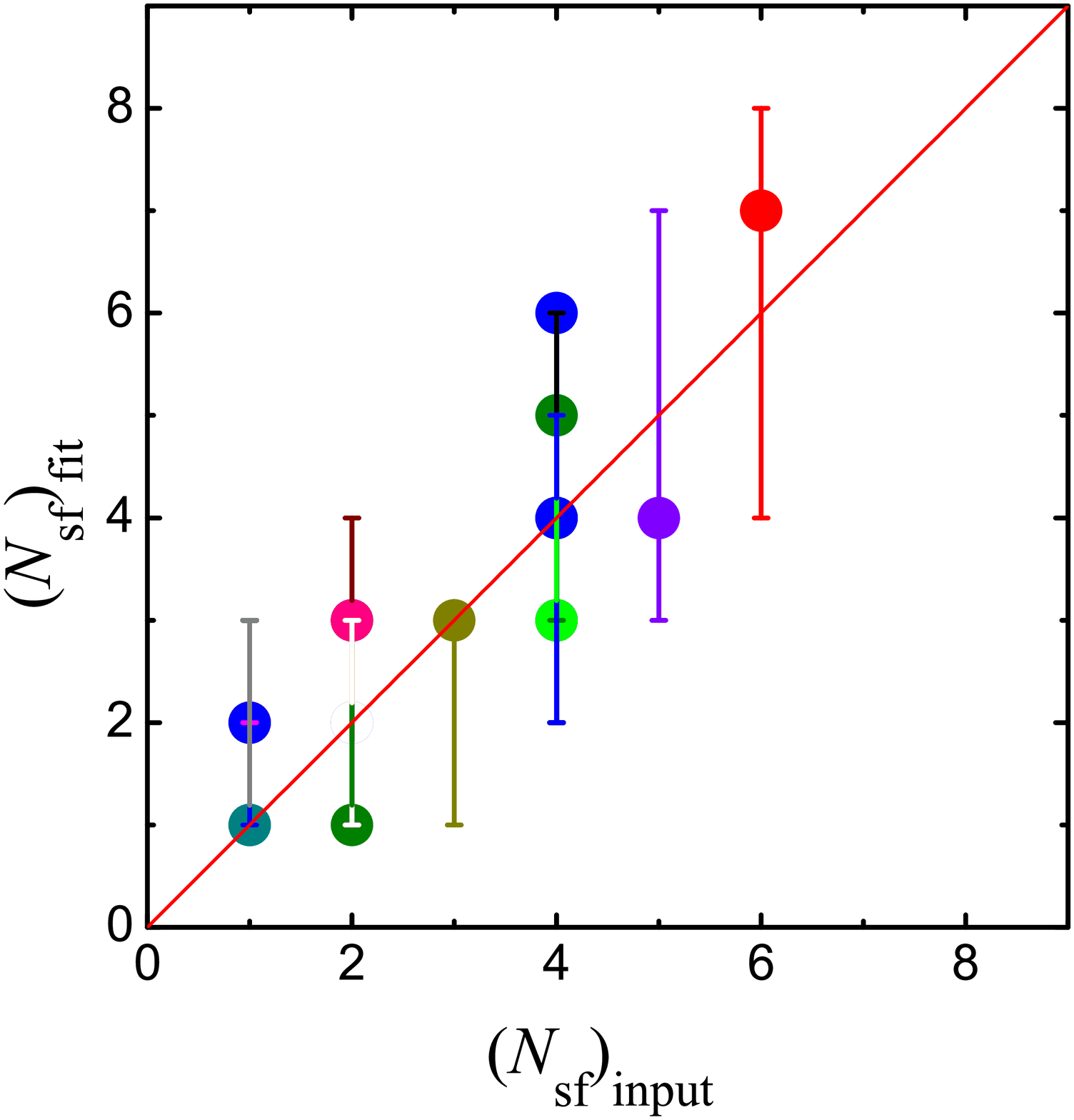}
\end{minipage}%
\begin{minipage}[t]{0.5\linewidth}
\centering
\includegraphics[width=0.7\textwidth]{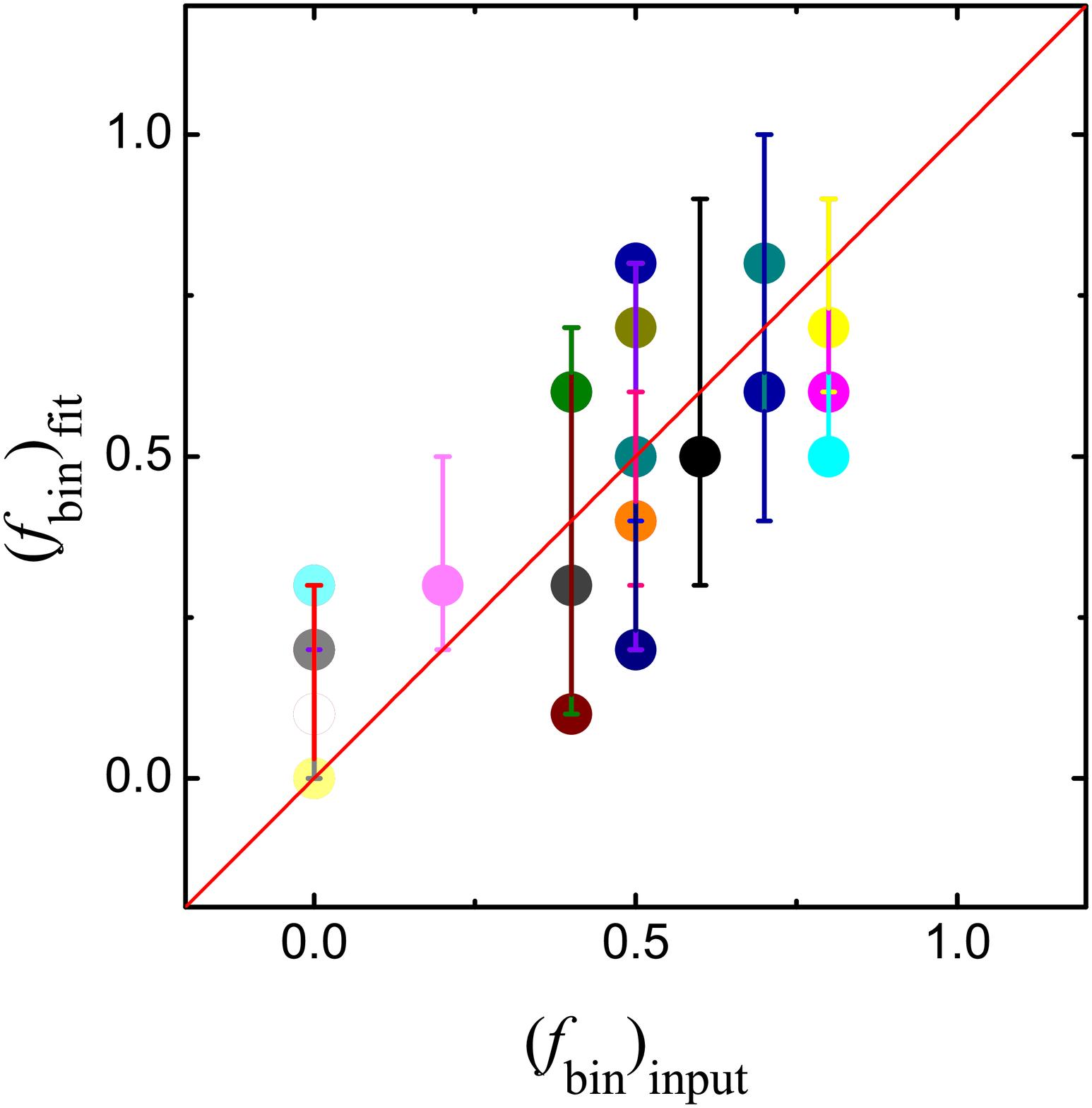}
\end{minipage}
\begin{minipage}[t]{0.5\linewidth}
\centering
\includegraphics[width=0.65\textwidth]{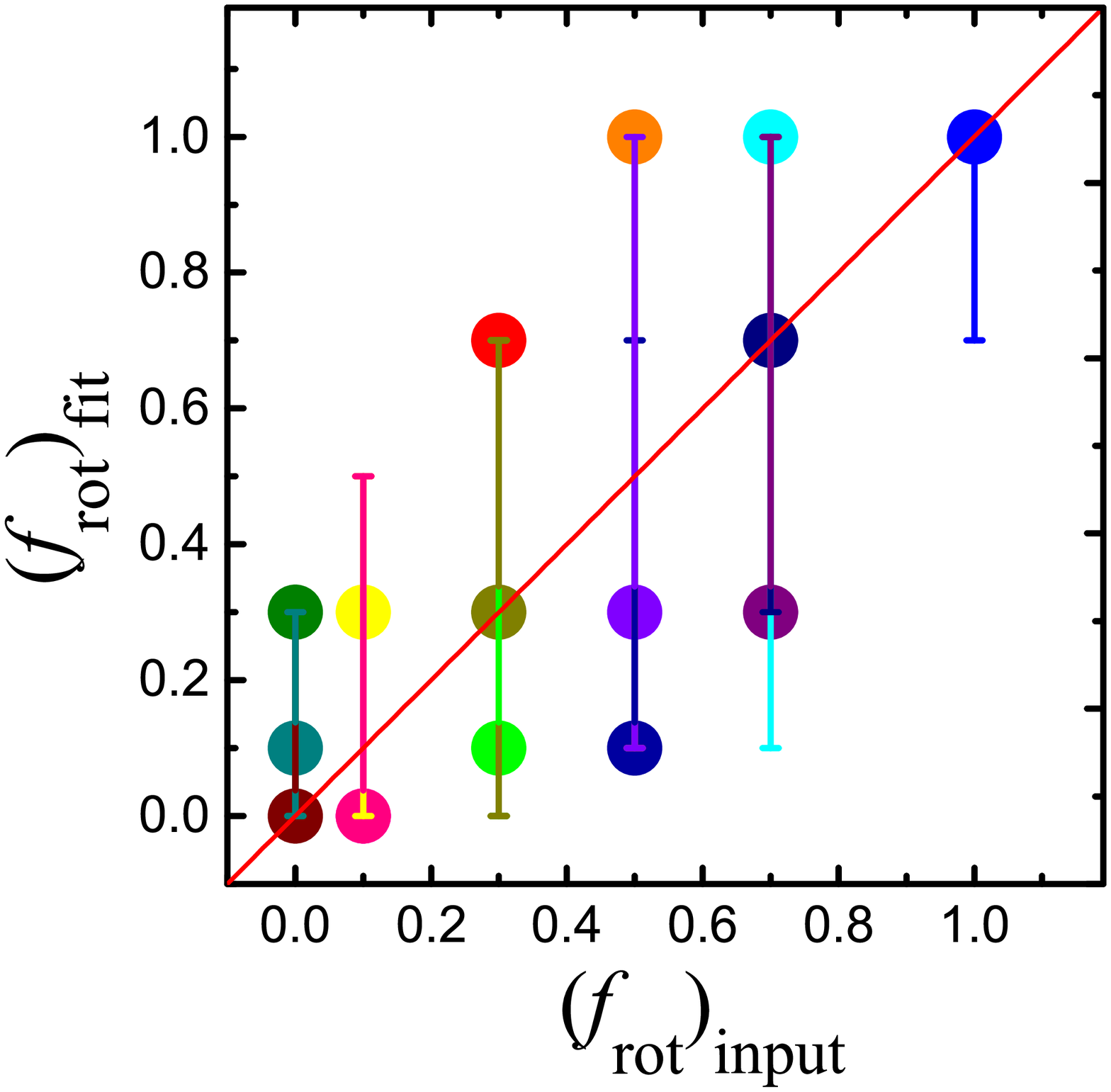}
\end{minipage}%
\caption{Comparison of input and fitted parameters of 51 simulated star clusters. $N_{\rm sf}$, $f_{\rm bin}$ and $f_{\rm rot}$ denote number of star burst with an interval of 0.1\,Gyr, binary fraction and rotating star fraction. Error bars indicate 1 $\sigma$ uncertainties.}
\end{figure}

\section{Application to four star clusters}
Four star clusters, i.e.,  NGC6362, NGC6652, NGC6838 and M67, are used for testing the new tool.
All of them do not have the presence of obviously extended main-sequence turn off and therefore can be fitted via SSPs.
Readers can check our previous paper, \cite{2015ApJ...802...44L}, for a detailed study of the CMD of NGC1651, which presents an extended main-sequence turn off.
There have been some studies about the test clusters, and this enables us to compare our results with previous results.
The data of M67 are directly taken from \cite{2008A&A...484..609Y}.
Those of three other clusters (i.e., NGC6362, NGC6652 and NGC6838) are obtained from the \emph{HST}
archive, which are observed with the Wide Field Planetary Camera 2 (WFPC2) between 1996 and 2000. Images were obtained using the F439W ($B$) and F555W ($V$) filters.
The exposure times of clusters NGC6362, NGC6652 and NGC6838 in F439W filter are 100, 100 and 160 seconds, and those in F555W filter are 40, 30 and 50 seconds, respectively.
We handle the \emph{HST} data using the stellar photometry package of \cite{2000PASP..112.1383D} (HSTphot),
because this package is specially designed for dealing with \emph{HST} WFPC2 images and it has been widely used.
Finally, we obtain the observed CMDs in $B$ and $V$ bands for three clusters.
The \emph{HST} magnitudes are transformed to $B$ and $V$ magnitudes by HSTphot.
Following some previous works, e.g., \cite{2010MNRAS.403.1156R}, magnitude uncertainties are estimated via an AST technique,
and the results are shown by Fig. 6.
Then Fig. 7 shows the observed CMDs (black points).
We can see the evolutionary structures, including main sequence, main-sequence turn off, Hertzsprung gap, and red giant, clearly.
Such CMDs are ideal for CMD studies.

When we use $Powerful~CMD$ to fit the CMDs of four clusters,
all observed CMDs are reproduced well.
The best-fit CMDs (red points) are compared to the observed ones via Fig. 7.
The results from stellar population models without binaries are the same as those with binaries, except colour excess.
The best-fit parameters are listed in Table 4 while 1 $\sigma$ ranges in Table 5.
``SSP-fit'' and ``CSP-fit'' denote the results from SSP and CSP models, respectively.
Note that in the fitting for M67, magnitude uncertainties are not considered as there is no available data.
Because some other works have studied these clusters, we compare our results with others in Table 4.
``Piotto'', ``Forbes'' and ``Yadav'' in Table 4 refer to the works of \cite{2002A&A...391..945P}, \cite{2010MNRAS.404.1203F} and \cite{2008A&A...484..609Y}, respectively.
It is shown that most of our results are in consistent with previous works, although different stellar population models and fitting methods are used.
In detail, the $(m-M)$ and age values obtained in this work are similar to other works, with only a small difference ($<$ 0.5\,mag and 0.4\,Gyr).
The metallicities of NGC6838 and M67 agrees well with previous works.
Although there are differences for the metallicities of NGC6362 and NGC6652, it is actually limited by the small number (eight) of metallicities of theoretical stellar populations.
If more metallicities are taken for theoretical populations, the results will be possibly closer.
Moreover, we find that $Powerful~CMD$ reports smaller colour excesses than previous results for all clusters.
This is reasonable, because binaries are taken into account by this work.
Some unresolved binaries are located at the right of main sequence, and this makes it able to fit the observed CMDs with smaller colour excesses.
We can see that a high binary fraction (0.8) is determined for NGC6362 and NGC6652.
The result is not strange, because binaries here mean those with orbital periods less than 100\,yr at zero age, rather than interacting binaries or main sequence binaries with large ($>$ 0.5) mass ratio. In other words, binaries in this paper includes all kinds of binaries. The binary components can be any type of stars, including black hole.
Thereby, the binary fractions derived by $Powerful~CMD$ usually seem larger than other works, e.g., \cite{milone2012}. The effect of binaries are studied widely, e.g., \cite{2009RAA.....9..191L,Yang2011,2014ApJ...789...88J}.
In addition, $Powerful~CMD$ reports an obviously smaller distance modulus for M67 compared to previous result.
This is also caused by the inclusion of binaries.
As we see in the last panel of Fig. 7, binaries enable us to fit the right part below turn-off with a smaller distance modulus
compared to the case of single-star populations.

In above test, we assume that these clusters are SSPs, because their CMDs seem similar to the isochrones of SSPs.
However, some of them are possibly CSPs, which was suggested by e.g., \cite{2015A&A...391..945P,milone2017}.
We therefore study the CMDs of star clusters NGC6362, NGC6652 and NGC6838 using CSP models.
We are finally shown that CSPs can fit the CMDs of three clusters better than SSPs.
Fig. 8 compares the observed and best-fit models for two clusters and Table 6 shows the best-fit results of three clusters.
We find that CSP models (Table 6) lead to smaller binary fractions comparing to SSP models (Table 4), while other parameters are similar.
The result is reasonable because both multiple population and stellar binarity contribute to stars in the region by blue box (Fig. 8).
In this case, the presence of multiple populations in star clusters affect the result from $Powerful~CMD$.

\begin{figure}
\begin{minipage}[t]{\linewidth}
\centering
\includegraphics[width=\textwidth]{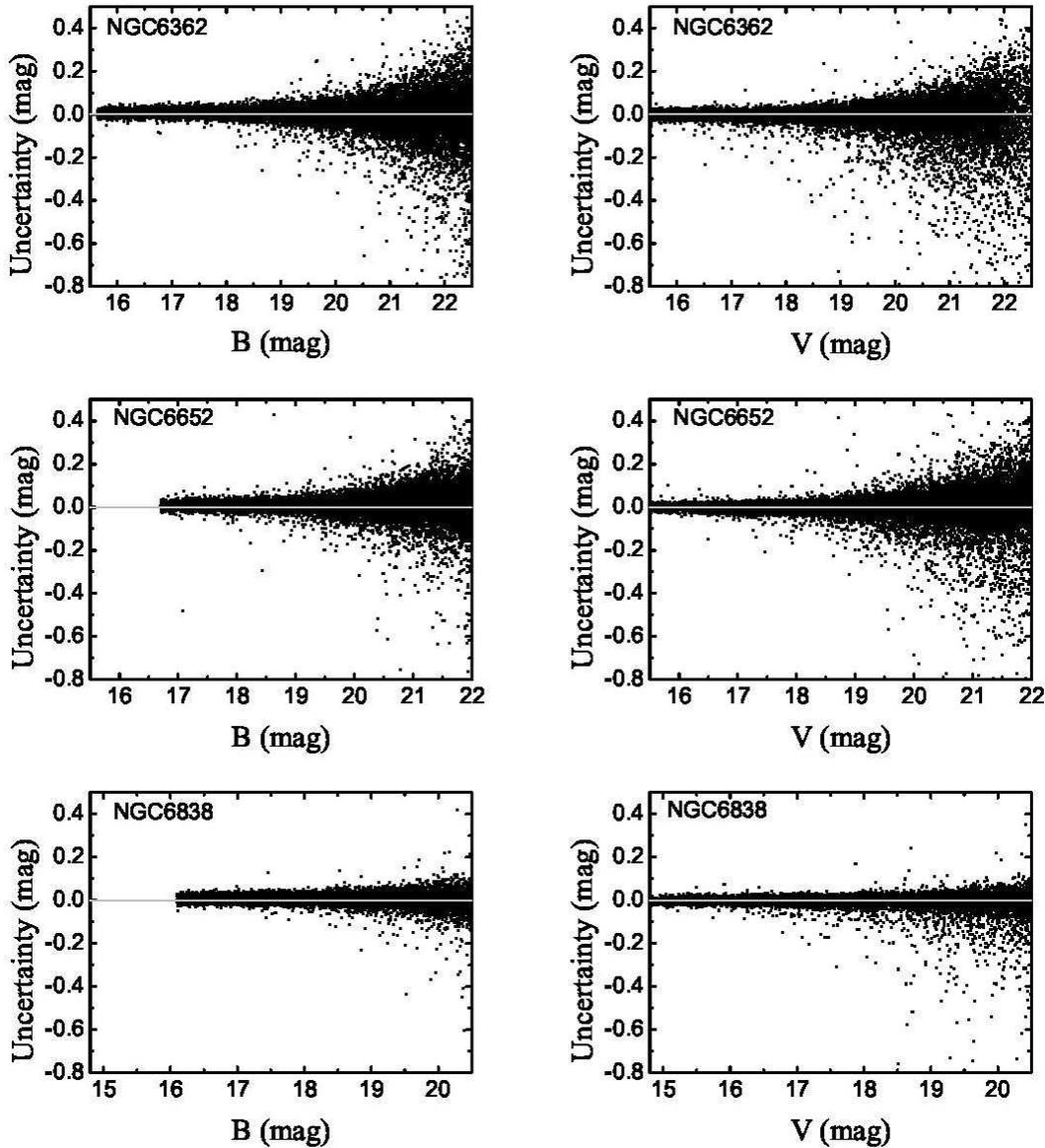}
\end{minipage}
  \caption{Magnitude uncertainty as a function of magnitude for clusters NGC6362, NGC6652, and NGC6838.}
\end{figure}

\begin{figure}
\begin{minipage}[t]{0.5\linewidth}
\centering
\includegraphics[width=\textwidth]{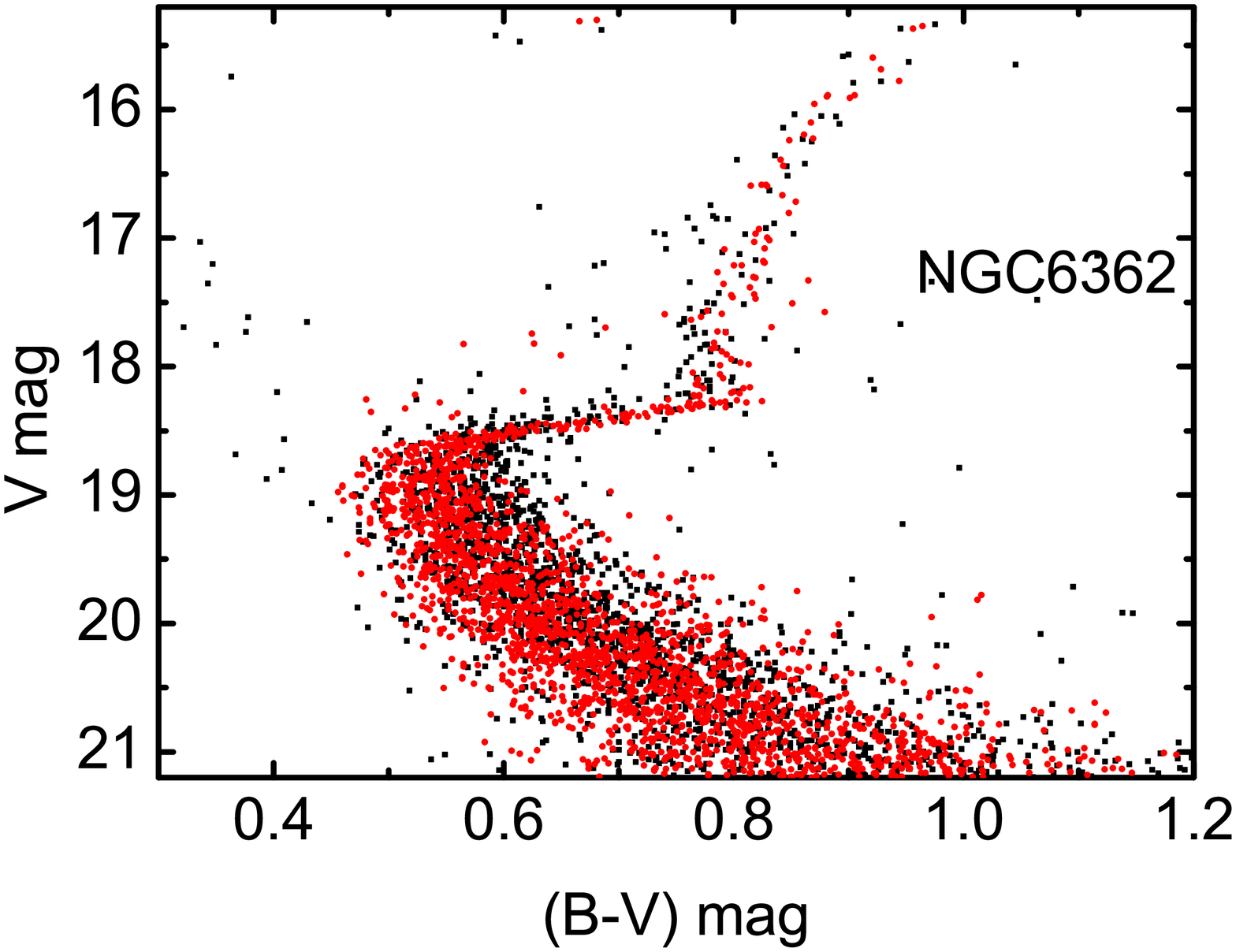}
\end{minipage}%
\begin{minipage}[t]{0.5\linewidth}
\centering
\includegraphics[width=\textwidth]{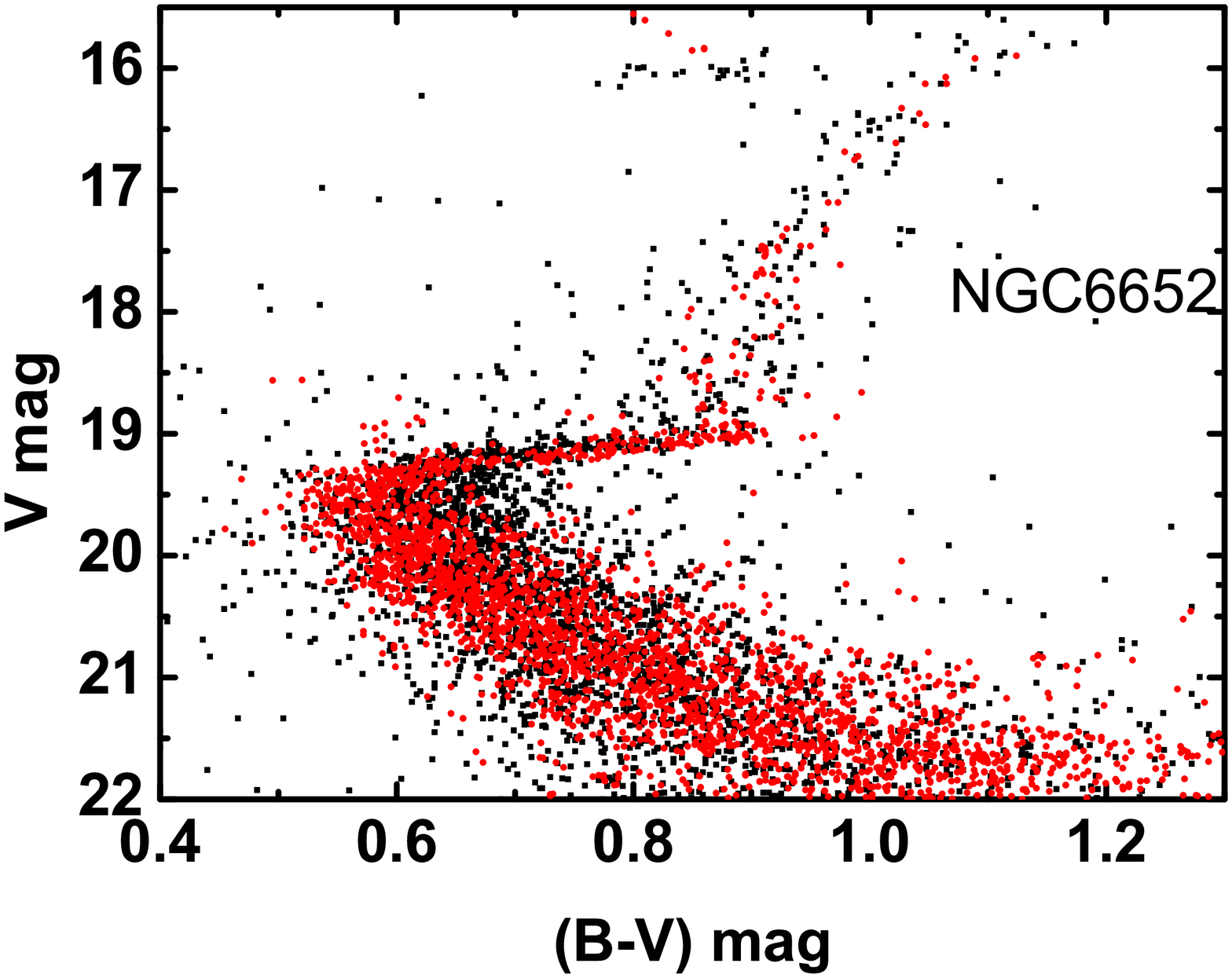}
\end{minipage}
\begin{minipage}[t]{0.5\linewidth}
\centering
\includegraphics[width=\textwidth]{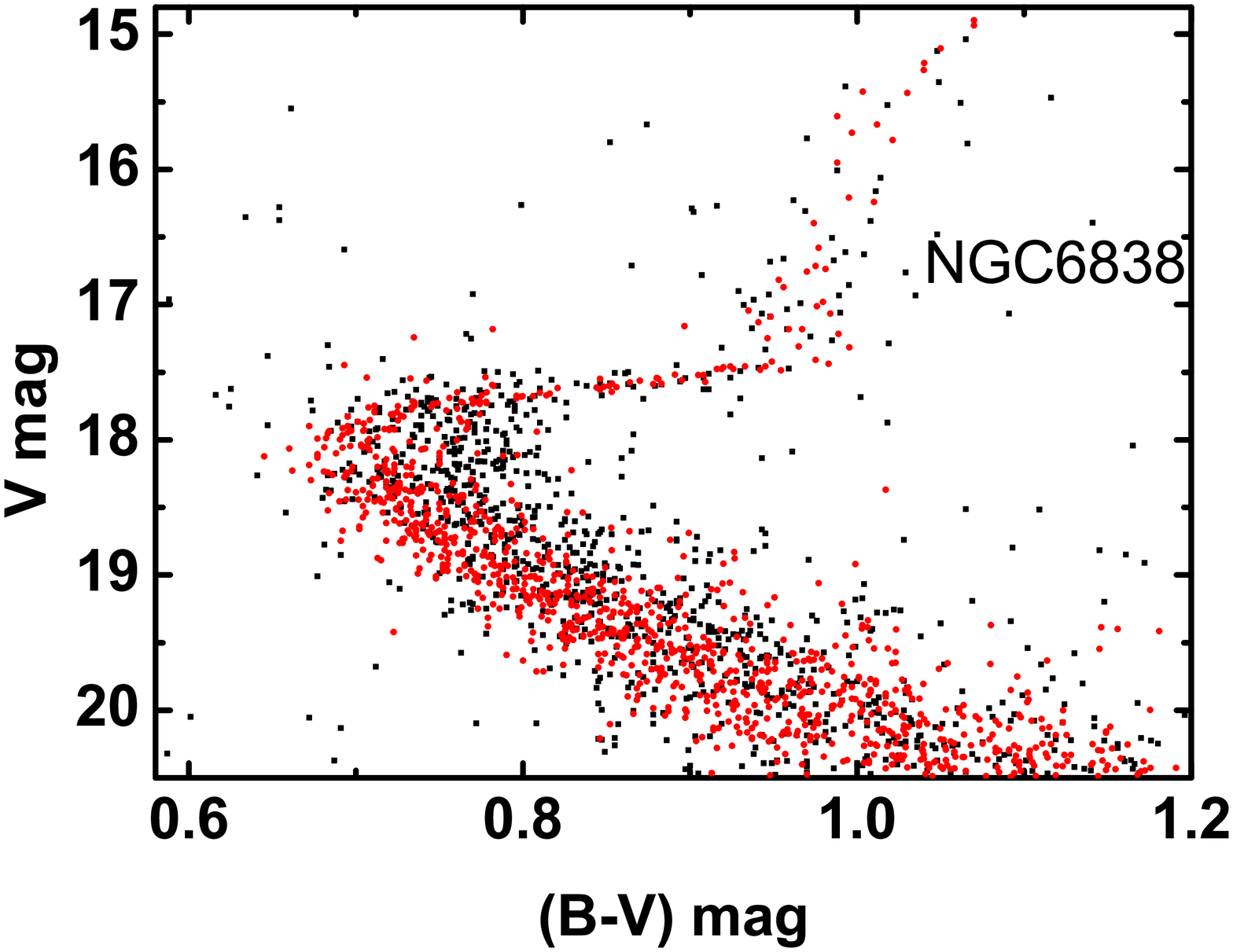}
\end{minipage}%
\begin{minipage}[t]{0.5\linewidth}
\centering
\includegraphics[width=\textwidth]{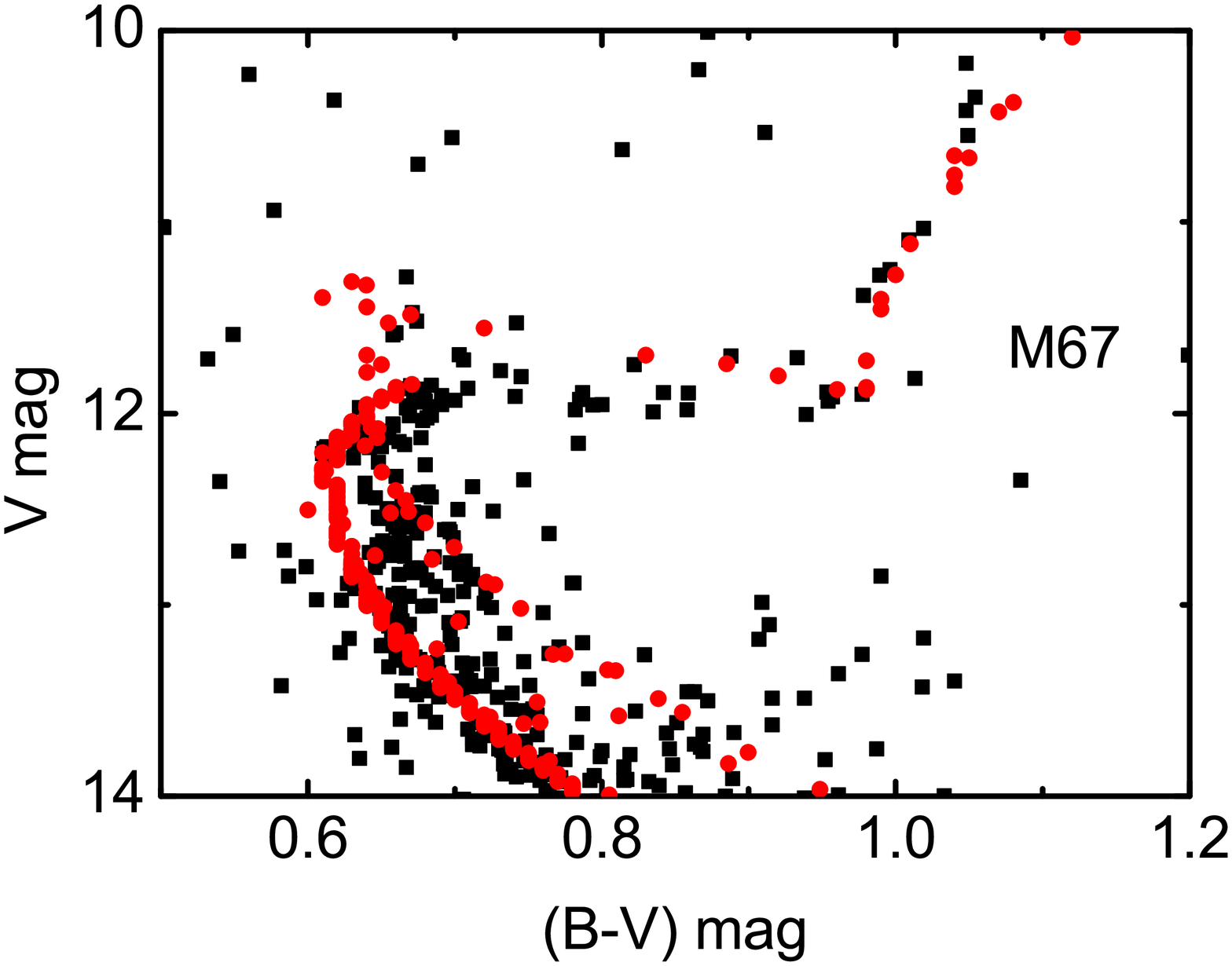}
\end{minipage}
  \caption{Comparison of observed (black) and SSP-fit (red) CMDs of clusters NGC6362, NGC6652, NGC6838 and M67.
  The observational uncertainties in magnitudes of M67 have not been considered.}
\end{figure}

\begin{figure}
\begin{minipage}[t]{0.5\linewidth}
\centering
\includegraphics[width=\textwidth,height=5cm] {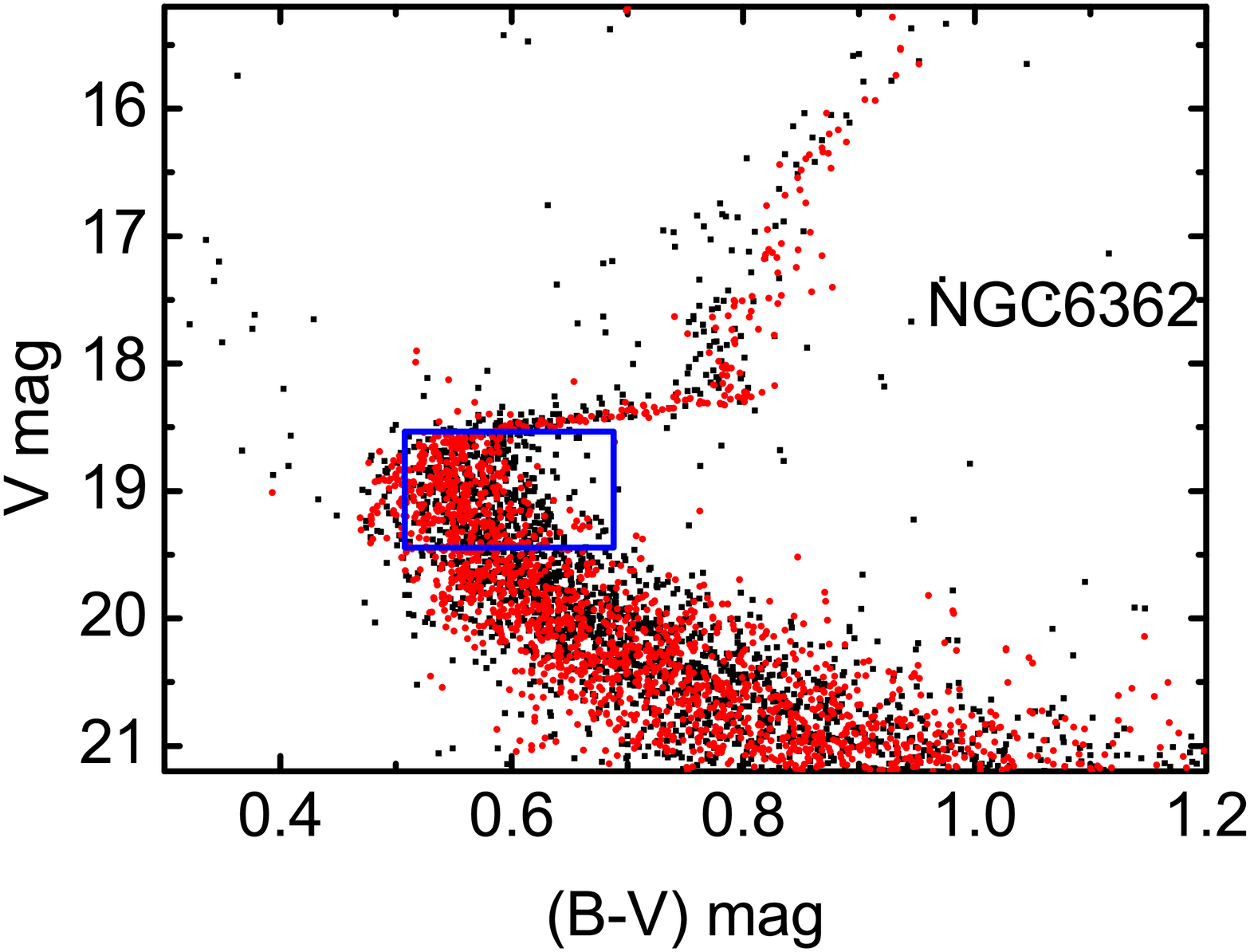}
\end{minipage}%
\begin{minipage}[t]{0.5\linewidth}
\centering
\includegraphics[width=\textwidth,height=5cm] {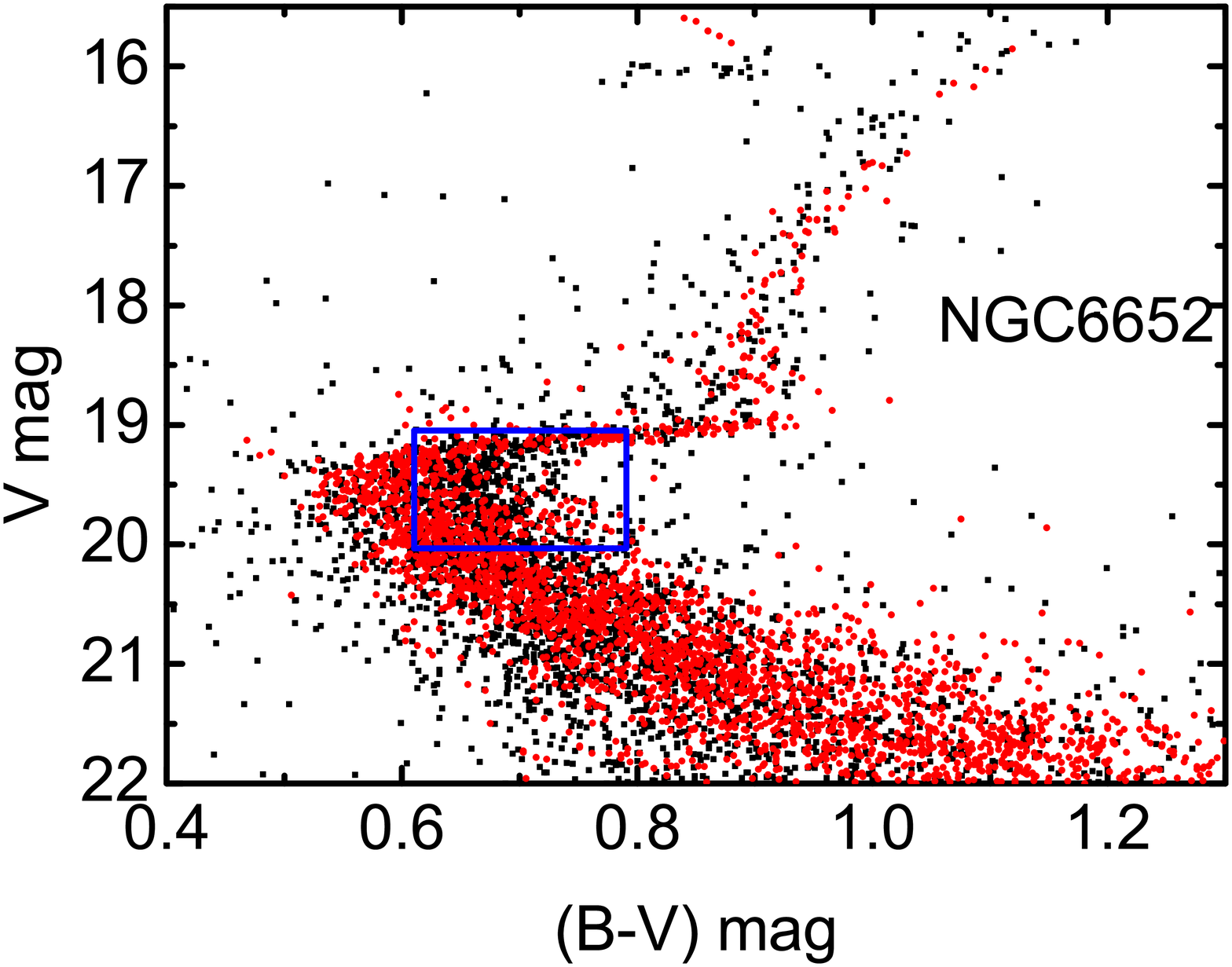}
\end{minipage}
  \caption{Comparison of observed (black) and CSP-fit (red) CMDs of clusters NGC6362 and NGC6652. Blue boxes show the regions that CSP models can reproduce better than SSP models. }
\end{figure}

  \begin{table*} 
 \caption{Best-fit parameters of star clusters NGC6362, NGC6652, NGC6838 and M67 from SSP models, together with other results.
 Sign apostrophe (') denotes the results from other works (see reference).
 Binary fraction $Fb$ are for all binaries with orbital period less than 100\,yr, rather than interacting binaries.}
 \label{symbols}
 \begin{tabular}{cccccccccccc}
  \hline\hline
Cluster	&$(m-M)$  &$(m-M)$'      &$E(B-V)$    &$E(B-V)$'     &$Z$    &$Z$'   &Age   &Age'     &$Fb$ & Reference     \\
        &[mag]    &[mag]         &[mag]   &[mag]     &	     &       &[Gyr] &[Gyr]            &                     \\
 \hline
NGC6362 &	14.41	    &	14.79	    &	0.01	     &	0.08	     &	0.004	&	0.0020, 0.0024	&	13.2	   &	 13.57	    &	0.8	&	Piotto, Forbes	\\				
NGC6652	&	15.13	    &	15.19	    &	0.04	     &	0.09	     &	0.008	&	0.0021, 0.0022	&	12.8	   &	 12.93	    &	0.8	&	Piotto, Forbes	\\				
NGC6838	&	14.22	    &	13.75	    &	0.20	     &	0.25	     &	0.004	&	0.0037	        &	13.4	   &	 13.70	    &	0.6	&	Piotto, Forbes	\\				
M67	    &	9.23	    &	9.56-9.72	&	0.10	     &	0.041	     &	0.020	&	0.0209-0.0219	&	3.6	       &	 3.5-4.8	    &	0.3	&	Yadav	\\		
 \hline
 \end{tabular}
 \end{table*}

 \begin{table*} 
 \caption{Ranges of SSP-fit parameters of star clusters NGC6362, NGC6652, NGC6838 and M67.
 The ranges correspond to 1 $\sigma$ confidence.}
 \label{symbols}
 \begin{tabular}{cccccccccccc}
  \hline\hline
Cluster	&$(m-M)$ range            &$E(B-V)$ range        &$Z$ range                &Age range               &$Fb$ range     \\
        &[mag]              &[mag]            &	                   &[Gyr]              &          \\
 \hline
NGC6362	 &	14.32--14.59	&	0.01--0.04    &     0.004          &	13.2--14.3     &	0.5--1.0\\
NGC6652	 &	15.00--15.40	&	0.01--0.08    &     0.008--0.010   &	11.8--13.8     &	0.6--0.9\\
NGC6838	 &	12.99--14.29	&	0.16--0.24    &     0.004          &	12.4--14.4     &	0.4--0.8\\
M67	     &	9.03--9.62	    &	0.07--0.19    &     0.010--0.030   &	3.1--4.1	   &	0.2--0.4\\
 \hline
 \end{tabular}
 \end{table*}

 \begin{table*} 
 \caption{Best-fit parameters of star clusters NGC6362, NGC6652 and NGC6838 from CSP models.}
 \label{symbols}
 \begin{tabular}{cccccccccccc}
  \hline\hline
Cluster	&$(m-M)$           &$E(B-V)$      &$Z$            &Ages             &$Fb$   &star formation mode \\
        &[mag]             &[mag]         &	              &[Gyr]            &       &            \\
 \hline
NGC6362   &14.42  &0.02    &0.004  &13.0--13.5    &0.7     &decreasing\\
NGC6652   &14.94  &0.03    &0.008  &12.6--12.7    &0.7     &homogeneous\\
NGC6838   &14.13  &0.21    &0.004  &13.1--13.5    &0.5     &homogeneous\\
 \hline
 \end{tabular}
 \end{table*}

\section{discussions and conclusions}
We present a new tool for CMD studies, $Powerful~CMD$, in this paper.
The new tool can be used for building theoretical CMDs of various kinds of stellar populations,
and for determining eight parameters of star clusters from CMDs.
The relevant stellar population synthesis model (i.e., ASPS),
building technique of synthetic CMDs, and CMD fitting method were introduced first.
Then we used $Powerful~CMD$ to build the CMDs of some artificial star clusters,
and check the efficiency of $Powerful~CMD$ using these simulated CMDs.
It is shown that the new tool has the ability to determine cluster parameters correctly.
Finally, the new tool was applied to four star clusters to determine their distance moduli,
colour excesses, metallicities, ages, and binary fractions.
The observed CMDs were fitted well, and the best-fit parameters agree with previous works as a whole,
although the inclusion of binaries in theoretical stellar population models leads to less colour excesses for three clusters.
This implies that $Powerful~CMD$ is a reliable tool for most CMD studies,
in particular for the studies with the $HST$ CMDs.

A limitation of the current version of $Powerful~CMD$ is that the star formation mode of about 20\% star clusters can not be automatically determined well.
This is caused by the degeneracy of different parameters.
$Powerful~CMD$ has supplied a function to determine the detailed star formation histories when other parameters are known.
If we can determine the stellar population types (SSP or CSP) via other methods, e.g., spectra,
$Powerful CMD$ will be able to determine the star formation histories.
In addition, this tool still contains some uncertainties,
which may result from the uncertainties in the modeling of stellar evolution (including single stars, binaries, rotators),
assumptions of stellar properties (e.g., IMF, distributions of binary separation and eccentricity, and distribution of stellar rotation rate),
estimation of observational uncertainties, and statistics for CMD fitting.
We will study them deeply in the future and improve the code.

Moreover, $Powerful~CMD$ utilizes a large size of data.
It makes not easy to spread this tool.
The authors will be glad to serve for all astronomers for free.
Meanwhile, we are trying to make the tool and data available to the public as soon as possible.

\normalem
\begin{acknowledgements}We thank the referee for useful comments.
This work is supported by the Chinese National Science Foundation (Grant No. 11563002, 11373003), Joint Research Project of Sino-German Center (GZ1284), National Key Basic Research Program of China (973 Program No. 2015CB857002),
and Yunnan Science Foundation (project name ``\emph{Study of Binary Fraction of Star Clusters}'').
We thank Prof. Li Chen of Shanghai Astronomical Observatory for suggestions.
\end{acknowledgements}


\begin{thebibliography}{42}
\providecommand\natexlab[1]{#1}
\providecommand\JournalTitle[1]{#1}

\bibitem[{Anderson} {et~al.} (2008)]{2008Anderson}
{Anderson}, J., {Sarajedini}, A., {Bedin}, L.~R., {King}, I.~R., {et~al.} 2008, \aj, 135, 2055

\bibitem[{Bertelli} {et~al.}(2003)]{2003AJ....125..770B}
{Bertelli}, G., {Nasi}, E., {Girardi}, L., {et~al.} 2003, \aj, 125, 770

\bibitem[{Brandt} \& {Huang}(2015)]{2015ApJ...807...25B}
{Brandt}, T.~D., \& {Huang}, C.~X. 2015, \apj, 807, 25

\bibitem[{Cignoni} \& {Shore}(2006)]{2006A&A...454..511C}
{Cignoni}, M., \& {Shore}, S.~N. 2006, \aap, 454, 511

\bibitem[{Da Rio} {et~al.}(2010)]{2010ApJ...723..166D}
{Da Rio}, N., {Gouliermis}, D.~A., \& {Gennaro}, M. 2010, \apj, 723, 166

\bibitem[{D'Antona} {et~al.}(2015)]{2015MNRAS.453.2637D}
{D'Antona}, F., {Di Criscienzo}, M., {Decressin}, T., {et~al.} 2015, \mnras,
  453, 2637

\bibitem[{De Gennaro} {et~al.}(2009)]{2009ApJ...696...12D}
{De Gennaro}, S., {von Hippel}, T., {Jefferys}, W.~H., {et~al.} 2009, \apj,
  696, 12

\bibitem[{Dolphin}(2000)]{2000PASP..112.1383D}
{Dolphin}, A.~E. 2000, \pasp, 112, 1383

\bibitem[{Dolphin}(2002)]{2002MNRAS.332...91D}
{Dolphin}, A.~E. 2002, \mnras, 332, 91

\bibitem[{Forbes} \& {Bridges}(2010)]{2010MNRAS.404.1203F}
{Forbes}, D.~A., \& {Bridges}, T. 2010, \mnras, 404, 1203

\bibitem[{Fusi Pecci} {et~al.}(1996)]{1996Fusi}
{Fusi Pecci}, F., {Buonanno}, R., {Cacciari}, C., {Corsi}, C.~E., {et~al.} 1996, \aj, 112, 1461

\bibitem[{Georgy} {et~al.}(2013)]{2013A&A...553A..24G}
{Georgy}, C., {Ekstr{\"o}m}, S., {Granada}, A., {et~al.} 2013, \aap, 553, A24

\bibitem[{Harris} \& {Zaritsky}(2001)]{2001ApJS..136...25H}
{Harris}, J., \& {Zaritsky}, D. 2001, \apjs, 136, 25

\bibitem[{Hurley} {et~al.}(2002)]{2002MNRAS.329..897H}
{Hurley}, J.~R., {Tout}, C.~A., \& {Pols}, O.~R. 2002, \mnras, 329, 897

\bibitem[{Hurley} \& {Tout}(1998)]{1998MNRAS.300..977H}
{Hurley}, J., \& {Tout}, C.~A. 1998, \mnras, 300, 977

\bibitem[{Jiang} {et~al.}(2014)]{2014ApJ...789...88J}
{Jiang}, D., {Han}, Z., \& {Li}, L. 2014, \apj, 789, 88

\bibitem[{Kerber} \& {Santiago}(2005)]{2005A&A...435...77K}
{Kerber}, L.~O., \& {Santiago}, B.~X. 2005, \aap, 435, 77

\bibitem[{Kerber} {et~al.}(2002)]{2002A&A...390..121K}
{Kerber}, L.~O., {Santiago}, B.~X., {Castro}, R., \& {Valls-Gabaud}, D. 2002,
  \aap, 390, 121

\bibitem[{Li} \& {Han}(2008{\natexlab{a}})]{2008MNRAS.387..105L}
{Li}, Z., \& {Han}, Z. 2008{\natexlab{a}}, \mnras, 387, 105

\bibitem[{Li} \& {Han}(2008{\natexlab{b}})]{2008ApJ...685..225L}
{Li}, Z., \& {Han}, Z. 2008{\natexlab{b}}, \apj, 685, 225

\bibitem[{Li} {et~al.}(2009)]{2009RAA.....9..191L}
{Li}, Z.-M., {Han}, Z.-W., {Li} 2009,
  Research in Astronomy and Astrophysics, 9, 191

\bibitem[{Li} {et~al.}(2010)]{2010RAA....10..135L}
{Li}, Z.-M., {Mao}, C.-Y., {Li}, R.-H., {Li}, R.-X., \& {Li}, M.-C. 2010,
  Research in Astronomy and Astrophysics, 10, 135

\bibitem[{Li} {et~al.}(2015)]{2015ApJ...802...44L}
{Li}, Z., {Mao}, C., \& {Chen}, L. 2015, \apj, 802, 44

\bibitem[{Li} {et~al.}(2012{\natexlab{a}})]{2012ApJ...761L..22L}
{Li}, Z., {Mao}, C., {Chen}, L., \& {Zhang}, Q. 2012{\natexlab{a}}, \apjl, 761,
  L22

\bibitem[{Li} {et~al.}(2013)]{2013ApJ...776...37L}
{Li}, Z., {Mao}, C., {Chen}, L., {Zhang}, Q., \& {Li}, M. 2013, \apj, 776, 37

\bibitem[{Li} {et~al.}(2016)]{2016arXiv1604.07156}
{Li}, Z., {Mao}, C., {Zhang}, L., {Zhang}, X., \& {Chen}, L. 2016, \apjs, 225, 7
\bibitem[{Li} {et~al.}(2012{\natexlab{b}})]{2012MNRAS.424..874L}
{Li}, Z., {Zhang}, L., \& {Liu}, J. 2012{\natexlab{b}}, \mnras, 424, 874

\bibitem[{Mackey \& Broby Nielsen}(2007)]{2007Mackey}
{Mackey}, A.~D., {Broby Nielsen}, P. 2007, \mnras, 379, 151

\bibitem[{Mackey} {et~al.}(2008)]{2008MNRAS.424..874L}
{Mackey}, A.~D. and {Broby Nielsen}, P. and {Ferguson}, A.~M.~N.,
\& {Richardson}, J.~C. 2008, \apjl, 681, L17

\bibitem[{Mieske} {et~al.}(2006)]{2006Mieske}
{Mieske}, S., {Jord{\'a}n}, A., {C{\^o}t{\'e}}, P., {Kissler-Patig}, {et~al.} 2006, \apj, 653, 193

\bibitem[{Milone} {et~al.}(2009)]{milone2009}
{Milone}, A.P., {Stetson}, P.B., {Piotto}, G., {Bedin}, L.R., \& et al., 2009, \aap, 503, 755

\bibitem[{Milone} {et~al.}(2012)]{milone2012}
{Milone}, A.P., {Piotto}, G., {Bedin}, L.R., {Aparicio}, A., \& et al., 2012, \aap, 540, A16

\bibitem[{Milone} {et~al.}(2017)]{milone2017}
{Milone}, A.~P., {Piotto}, G., {Renzini}, A., {Marino}, A.~F. 2017, \mnras, 464, 3636

\bibitem[{Naylor} \& {Jeffries}(2006)]{2006MNRAS.373.1251N}
{Naylor}, T., \& {Jeffries}, R.~D. 2006, \mnras, 373, 1251

\bibitem[{Niederhofer} {et~al.}(2016)]{2016A&A...586A.148N}
{Niederhofer}, F., {Bastian}, N., {Kozhurina-Platais}, V., {et~al.} 2016, \aap,
  586, A148

\bibitem[{Niederhofer} {et~al.}(2015)]{2015MNRAS.453.2070N}
{Niederhofer}, F., {Georgy}, C., {Bastian}, N., \& {Ekstr{\"o}m}, S. 2015,
  \mnras, 453, 2070

\bibitem[{Olsen} {et~al.}(1998)]{1998Olsen}
{Olsen}, K.~A.~G., {Hodge}, P.~W., {Mateo}, M., {Olszewski}, E.~W., {et~al.} 1998, \mnras, 300, 665

\bibitem[{Piotto} {et~al.}(2002)]{2002A&A...391..945P}
{Piotto}, G., {King}, I.~R., {Djorgovski}, S.~G., {et~al.} 2002, \aap, 391, 945

\bibitem[{Piotto} {et~al.}(2015)]{2015A&A...391..945P}
{Piotto}, G., {Milone}, A.~P., {Bedin}, L.~R., {Anderson}, J., {et~al.} 2015, \aj, 149, 91	


\bibitem[{Royer} {et~al.}(2007)]{2007A&A...463..671R}
{Royer}, F., {Zorec}, J., \& {G{\'o}mez}, A.~E. 2007, \aap, 463, 671

\bibitem[{Rubele} {et~al.}(2010)]{2010MNRAS.403.1156R}
{Rubele}, S., {Kerber}, L., \& {Girardi}, L. 2010, \mnras, 403, 1156

\bibitem[{Salpeter}(1955)]{1955ApJ...121..161S}
{Salpeter}, E.~E. 1955, \apj, 121, 161

\bibitem[{Sandquist} {et~al.} (1996)]{1996Sandquist}
{Sandquist}, E.~L., {Bolte}, M., {Stetson}, P.~B., {Hesser}, J.~E. 1996, \apj, 470, 910

\bibitem[{VandenBerg} {et~al.}(2013)]{2013VandenBerg}
{VandenBerg}, D.~A., {Brogaard}, K., {Leaman}, R., {Casagrande}, L. 2013, \apj, 775, 134

\bibitem[{von Hippel} {et~al.}(2006)]{2006ApJ...645.1436V}
{von Hippel}, T., {Jefferys}, W.~H., {Scott}, J., {et~al.} 2006, \apj, 645,
  1436

\bibitem[{Yadav} {et~al.}(2008)]{2008A&A...484..609Y}
{Yadav}, R.~K.~S., {Bedin}, L.~R., {Piotto}, G., {et~al.} 2008, \aap, 484, 609

\bibitem[{Yang} {et~al.}(2011)]{Yang2011}
{Yang}, W., {Meng}, X., {Bi}, S., {Tian}, Z., {et~al.} 2011, \apjl, 731, 37

\bibitem[{Yang} {et~al.}(2013)]{Yang2013}
{Yang}, W., {Bi}, S., {Meng}, X., {Liu}, Z. 2013, \apj, 776, 112

\bibitem[{Zhongmu}(2011)]{2011ASPC..451...51Z}
{Zhongmu}, L. 2011, in Astronomical Society of the Pacific Conference Series,
  Vol. 451, 9th Pacific Rim Conference on Stellar Astrophysics, ed. S.~{Qain},
  K.~{Leung}, L.~{Zhu}, \& S.~{Kwok}, 51
\end{thebibliography}

\end{document}